\documentclass[12pt,preprint]{aastex}
 
\shortauthors{Sil'chenko et al.}
\shorttitle{NGC 3368, NGC 3379, and NGC 3384}

\begin{document}

\title{The Leo I Cloud: Secular nuclear evolution of NGC 3379, NGC 3384,
           and NGC 3368?}
\slugcomment{Partly based on observations collected with the 6m
telescope (BTA) at the Special Astrophysical Observatory (SAO) of the
Russian Academy of Sciences (RAS).}

\author{O. K. Sil'chenko\altaffilmark{1}}
\affil{Sternberg Astronomical Institute, Moscow, 119992 Russia\\
       and Isaac Newton Institute of Chile, Moscow Branch\\
     Electronic mail: olga@sai.msu.su}

\altaffiltext{1}{Guest Investigator of the UK Astronomy Data Centre}

%\and

\author{A. V. Moiseev, V. L. Afanasiev}
\affil{Special Astrophysical Observatory, Nizhnij Arkhyz,
    369167 Russia\\
    Electronic mails: moisav@sao.ru, vafan@sao.ru}

%\and

\author{V. H. Chavushyan, J. R. Valdes}
\affil{Instituto Nacional de Astrof\'{\i}sica,
  Optica, y Electr\'onica, A.P. 51 y 216, C.P. 72000 Puebla, Pue., M\'exico\\
     Electronic mails: vahram@inaoep.mx, jvaldes@inaoep.mx}

\begin{abstract}

The central regions of the three brightest members of the Leo~I galaxy
group -- NGC~3368, NGC~3379, and NGC~3384 -- are investigated by means
of 2D spectroscopy. In all three galaxies we have found separate
circumnuclear stellar and gaseous subsystems -- more probably, disks --
whose spatial orientations and spins are connected to the spatial
orientation of the supergiant intergalactic H~I ring reported previously
by Schneider et al. (1983) and Schneider (1985, 1989). In NGC~3368 the
global gaseous disk seems also to be inclined to the symmetry plane of
the stellar body, being probably of external origin. Although
the rather young mean stellar age and spatial orientations of the
circumnuclear disks in NGC~3379, NGC~3384, and NGC~3368 could imply their
recent formation from material of the intergalactic H~I cloud,
the time scale of these secondary formation events, of order 3~Gyr,
does not support the collision scenario of Rood \&\ Williams (1985),
but is rather in line with the ideas of Schneider (1985, 1989) regarding
tidal interactions of the galaxies with the H~I cloud on timescales of
the intergroup orbital motions.

\end{abstract}


\section{Introduction}

The origin of S0 galaxies is long-standing problem which was posed
almost at the moment of the birth of the Hubble morphological
classification scheme. However,
this has been one of those rare cases when the first idea was correct,
this being
later confirmed more than once. From the beginning many investigators
thought that lenticulars formed (more exactly, transformed) from spirals.
Now with the advent of high-resolution imaging, including the
HST as well as some ground-based work, using adaptive optics,
there is direct evidence that rich clusters, presently
populated mostly by S0's, at $z=0.5 - 0.8$ contain a lot of spirals
which are
in the course being accreted by these clusters. Evidently, dense
environments, or deep potential wells, or hot intracluster gas
pressure, provides the conditions to transform the spirals
into lenticulars. Two famous theoretical papers concerning particular
mechanisms of this transformation must be mentioned here: \citet{spitbaad}
propose collisions between spiral galaxies, which are frequent in
dense clusters with high velocity dispersions, in which in a
passage of one galaxy through another, all the gas is swept out of
their galactic disks. The other paper is by \citet{ltcs0} who analysed a less
violent event during which tidal stripping removes a diffuse gaseous halo from
a spiral galaxy. The secular disk building by gas accretion onto the
equatorial plane having ceased, the gas remaining in the disk is
typically consumed by star formation in a few billion years.

Therefore one can easily imagine -- and even directly observe -- how
spirals transform into lenticulars in clusters. But clusters are not
the only place where S0's live; there are a lot of lenticulars in the
field and in loose groups. What is their origin? There are suggestions
that field S0s form in a different way from that of cluster S0s.
There exists one loose galaxy group where the scenario of
\citet{spitbaad} may apply: it is the Leo~I group. Twenty
years ago a unique supergiant intergalactic HI cloud was discovered
in this galaxy group \citep{leoh1_1}, and \citet{roodwil}
suggested that this gas was swept out of galactic disks during a collision
between NGC~3368 and NGC~3384 some $5 \times 10^8$ yrs ago. Now the
HI cloud is located exactly half-way between two galaxies, and NGC~3384
looks like a bona-fide lenticular galaxy. But Schneider himself
\citep{leoh1_2,leoh1_3} gave another explanation to this phenomenon.
The cloud has fainter filaments which encircle the galaxy pair
NGC 3384/NGC 3379; the whole HI-complex may be treated as a clumpy
gaseous ring with a radius of some 100 kpc. Velocities measured in
several clumps imply a Keplerian rotation of the ring with a period
of 4 Gyr, the SW-part of the ring being receding. If the intrinsic
shape of the ring is circular, it is inclined to the line of sight
with the line of nodes at $PA\approx 40^{\circ}$; interestingly,
the close orientation and sense of rotation is repeated by the global
stellar disk of NGC~3384. Even though symmetry (rotation) axes of the rest
of the galaxies of the group are not aligned with those of the HI-ring and
NGC~3384, \citet{leoh1_2,leoh1_3} conclude that this supergiant,
$1.7 \cdot 10^9$ $M_{\odot}$, HI-cloud may represent left-over primordial
(pre-galactic) gas from which all the galaxies of the group have been
formed.

As the HI-ring is rather massive, it contributes to the common
potential; in particular, tidal interactions of the group galaxies
with the ring are possible. \citet{leoh1_3} have analysed the possible
interaction of the ring with NGC~3368 -- the nearest and the brightest
neighbor of the NGC 3384/NGC 3379/HI-ring complex. He suggested
that the intergroup orbit of  NGC~3368 is inclined to the plane of the ring
and that this galaxy has had two close passages near the ring during a
Hubble time. In this case, the cycle time of interaction is about 5 Gyr,
instead of $5 \cdot 10^8$ yrs as proposed in the model of \citet{roodwil}.

Well after the first simulations of tidal interactions of galaxies
by \citet{toomre}, demonstrating spectacular outer tidal
structures (tails, bridges, etc.), it was realized that an external tidal
impulse affects also the innermost (circumnuclear) structure of the
galaxy -- see e.g. \citet{noguchi}. So a history of nuclear star formation
may reflect in some way the history of galaxy interactions, especially
if the galaxy possesses an early morphological type and lacks its own gas.
In this paper we study the properties of nuclear stellar populations in
NGC~3379, NGC~3384, and NGC~3368 -- three of the four brightest galaxies
of the Leo~I group and three nearest neighbors of the unique intergalactic
HI-ring. We are searching for signs of synchronous evolution of the nuclear
stellar populations in these galaxies; obviously, the characteristic time
of this evolution may help to discriminate between various scenarios
of the origin of the HI-ring.

The global parameters of the galaxies are given in Table~1.
The distance to the spiral galaxy NGC~3368 is determined from
Cepheid observations \citep{distceph}, and the distance to
the high surface brightness elliptical NGC~3379 by the
surface-brightness fluctuations method \citep{morshank};
the distance to NGC~3384 was assumed to be the same as to NGC~3379
as their planetary nebula systems seem to imply \citep{pndist}.

The layout of the paper is as follows. We report our observations and
other data which we use in this paper in Section~2. The radial
variations of the properties of the
stellar population are analysed in Section~3, and
in Section~4 two-dimensional velocity fields obtained
by means of  2D spectroscopy are presented. Section~5 presents a
discussion and our conclusions.

\begin{table}
\caption[ ] {Global parameters of the galaxies}
% %\begin{center}
\begin{flushleft}
\begin{tabular}{lccc}
\hline\noalign{\smallskip}
% % & & & \\
NGC & 3368 & 3379 & 3384 \\
\hline\noalign{\smallskip}
Type (NED$^1$) & SAB(rs)ab & E1 & SB(s)0-:  \\
$R_{25}$, kpc (LEDA$^2$) & 12.6 & 9.8 & 9.8 \\
$B_T^0$ (RC3$^3$) & 9.80 & 10.18 & 10.75  \\
$M_B$ (LEDA)  & --20.9 & --20.5 & --19.6 \\
$(B-V)_T^0$ (RC3) & 0.79 & 0.94 & 0.91 \\
$(U-B)_T^0$ (RC3) & 0.25 & 0.52 & 0.43 \\
$V_r$, $\mbox{km}\cdot \mbox{s}^{-1}$ (NED) & 897 &
 911  & 704  \\
Distance, Mpc  & 11.2 & 12.6 & 12.6 \\
Inclination (LEDA) & $55^{\circ}$ & $32^{\circ}$ & $90^{\circ}$  \\
{\it PA}$_{phot}$ (LEDA) & $5^{\circ}$ & $71^{\circ}$ & $53^{\circ}$  \\
\hline
\multicolumn{4}{l}{$^1$\rule{0pt}{11pt}\footnotesize
NASA/IPAC Extragalactic Database}\\
\multicolumn{4}{l}{$^2$\rule{0pt}{11pt}\footnotesize
Lyon-Meudon Extragalactic Database}\\
\multicolumn{4}{l}{$^3$\rule{0pt}{11pt}\footnotesize
Third Reference Catalogue of Bright Galaxies}
\end{tabular}
\end{flushleft}
\end{table}

\section{Observations and data reduction}

The spectral data which we analyse in this work were obtained with
two different integral-field spectrographs. Integral-field spectroscopy
is a rather new approach that was first proposed by Prof. G. Courtes
some 15 years ago -- for a description of the instrumental idea see
e.g. \citet{betal95}. It allows one to obtain simultaneously a set of
spectra in a wide spectral range from an extended area on the sky,
for example, from a central part of a galaxy. A 2D array of microlenses
provides a set of micropupils which are projected onto to the entrance
window of
a spectrograph. After reducing the full set of spectra
corresponding to the individual spatial elements, one obtains a list of
fluxes for the continuum and for the lines,
line-of-sight velocities, both for stars and ionized gas, and
of emission- and absorption-line equivalent widths which are usually
expressed as
indices in the well-formulated Lick system \citep{woretal}.
This list can be transformed into two-dimensional maps of the
above mentioned characteristics for the central part of the galaxy
under study. Besides the panoramic view benefits, such an approach
gives a unique opportunity to overlay various 2D distributions
over each other without any difficulties with positioning.
In this work we use the data from two 2D spectrographs: the fiber-lens
Multi-Pupil Fiber Spectrograph (MPFS) at the 6m telescope BTA of the
Special Astrophysical Observatory of the Russian Academy of Sciences
(SAO RAS) and the international Tigre-mode SAURON at the 4.2m William
Herschel Telescope at La Palma.

A new variant of the MPFS became operational
at the prime focus of the 6m telescope in 1998
(http://www.sao.ru/$\sim$gafan/devices/mpfs/).
With respect to the previous one, in the new variant
the field of view is increased and the common spectral range
is larger due to using fibers: they transmit light from $16\times 15$ square
elements of the galaxy image to the slit of the spectrograph together
with additional 16 fibers which transmit the sky background light
taken away from the galaxy, so separate sky exposures are not
necessary anymore. The size of one spatial element is approximately
$1\arcsec \times 1\arcsec$; a CCD TK $1024 \times 1024$ detector
is used. The reciprocal dispersion is 1.35~\AA\ per pixel, with
a rather stable spectral resolution of 5~\AA.
To calibrate the wavelength scale, we expose separately a spectrum
of a hollow cathode lamp filled with helium, neon, and argon;
an internal accuracy of linearizing the spectrum was typically
0.25~\AA\ in the
green and 0.1~\AA\ in the red. Additionally we checked the accuracy
and absence of a systematic velocity shift by measuring strong emission
lines of the night sky, [OI]$\lambda$5577 and [OI]$\lambda$6300.
We obtained the MPFS data mostly in two spectral ranges,
the green one, 4300--5600~\AA, and the red one, 5900--7200~\AA.
The green spectra are used to calculate
the Lick indices H$\beta$, Mgb, Fe5270, and Fe5335 which are suitable
to determine the mean (luminosity-averaged) metallicity, age, and Mg/Fe ratio
of old stellar populations \citep{worth94}. Also, they are used for
cross-correlating with a spectrum
of a template star, usually of K0III--K3III spectral type, to obtain
in such a way a line-of-sight velocity field for the stellar component
and a map of stellar velocity dispersion. The red spectral range
contains strong emission lines of H$\alpha$ and [NII]$\lambda$6583
and is used to derive line-of-sight velocity fields of the ionized gas.
To calibrate the new MPFS index system onto the standard Lick one,
we observed 15 stars from the list of \citet{woretal} during
four observing runs and calculated the
linear regression formulas to transform our index measurements to the
Lick system; the rms scatter of points near the linear regime
is about 0.2~\AA\ for all 4 indices under consideration, i.e.,
within the observational errors quoted by \citet{woretal}. To
correct the index measurements for the stellar velocity dispersion
which is usually substantially non-zero in the centers of early-type
galaxies, we smoothed the spectrum of the standard
star HD~97907 by a set of Gaussians of various widths; the derived
dependence of index corrections on $\sigma$ were approximated by
4th order polynomials and applied to the measured index values
before their calibration to the Lick system.

The second 2D spectrograph which data we use in this work is a new
instrument which is being operated at the 4.2m William Herschel
Telescope on La Palma
since 1999, named SAURON -- for its detailed description see
\citet{betal01}. We have taken the data for NGC~3379 and NGC~3384
from the open ING Archive of the UK Astronomy Data Centre. Briefly,
the field of view of this instrument is $41\arcsec \times 33\arcsec$
with spatial element sizes of $0\farcs 94 \times 0\farcs 94$.
The sky background at 2 arcminutes from the center of the galaxy
is exposed simultaneously with the target. The fixed spectral range is
4800-5400~\AA, the reciprocal dispersion is 1.11~\AA\--1.21~\AA\ per pixel
varying from the left to the right edge of the frame. The comparison
spectrum is neon, and the linearization is made using a 2nd order polynomial
with an accuracy of 0.07~\AA. The index system is
checked by using stars from the list of \citet{woretal} that have been
observed during the same observing runs as the galaxies. The
regressions describing the index system of the February-1999 run
when NGC~3379 was observed can be found in our paper
\citep{afsil02b}, and
the regressions for the March/April-2000 run are presented in Fig.~1.
The relations between instrumental and standard-system indices
were approximated by linear fits which were applied to
our measurements to calibrate them on to the standard Lick system.
The stellar velocity dispersion effect was corrected in the
same manner as for the MPFS data.

The full list of exposures made for NGC~3368, NGC~3379,
and NGC~3384 with two 2D spectrographs is given in Table~2.

\begin{table}
\footnotesize{\tiny}
\caption[ ] {2D spectroscopy of the galaxies studied}
% \begin{center}
\begin{flushleft}
\begin{tabular}{lllllcc}
\hline\noalign{\smallskip}
Date & Galaxy & Exposure & Configuration & Field
& Spectral range & Seeing \\
\hline\noalign{\smallskip}
20 Feb 99 & NGC~3379 & 60 min & WHT/SAURON+CCD $2k\times 4k$ &
$33\arcsec\times 41\arcsec$ & 4800-5400~\AA\ & $1\farcs 5$ \\
8 Feb 00 & NGC~3368 & 45 min & BTA/MPFS+CCD $1024 \times 1024$ &
$16\arcsec \times 15\arcsec $ & 4200-5600~\AA\ & $1\farcs 4$ \\
13 Feb 00 & NGC~3368 & 40 min & BTA/MPFS+CCD $1024 \times 1024 $ &
$16\arcsec \times 15\arcsec $ & 6000-7200~\AA\ & $2\farcs 3$ \\
28 Mar 00 & NGC~3368 & 140 min & BTA/MPFS+CCD $1024 \times 1024 $ &
$16\arcsec \times 15\arcsec $ & 4840-6210~\AA\ & $2\farcs 5$ \\
11 Dec 99 & NGC~3384 & 45 min & BTA/MPFS+CCD $1024 \times 1024$ &
$16\arcsec \times 15\arcsec $ & 4200-5600~\AA\ & $1\farcs 6$ \\
4 Apr 00 & NGC~3384 & 120 min & WHT/SAURON+CCD $2k\times 4k$ &
 $33\arcsec\times 41\arcsec$ & 4800-5400~\AA\ & $3\farcs 4$ \\
\hline
\end{tabular}
% \end{center}
\end{flushleft}
\end{table}

\begin{table}
\footnotesize{\tiny} \caption[ ]{IFP observations of NGC~3368 at the 6m telescope}
% \begin{center}
\begin{flushleft}
\begin{tabular}{lllc}
\hline\noalign{\smallskip} Date &  Exposure &  Spectral range & Seeing \\
\hline\noalign{\smallskip} 28 Feb 00 & $32\times150$ sec & around $H_\alpha$  &  2\farcs7\\
 28 Feb 00 & $32\times200$ sec & around [NII]$\lambda6583$  &  3\farcs5\\
\hline
\end{tabular}
% \end{center}
\end{flushleft}
\end{table}

\begin{table}
\caption[ ] {Photometric observations of the galaxies studied}
% \begin{center}
\begin{flushleft}
\begin{tabular}{cccccc}
\hline\noalign{\smallskip}
Date & Galaxy & Filter & Exposure time & Scale,$\arcsec$ per px & Seeing \\
\hline\noalign{\smallskip}
18 Apr 00 & NGC~3368 &  $J$ & 15 min & 0.3 & $1\farcs 5$ \\
18 Apr 00 & NGC~3368 &  $H$ & 12 min & 0.3 & $1\farcs 5$ \\
18 Apr 00 & NGC~3368 &  $K'$ & 12 min & 0.3 & $1\farcs 5$ \\
12 Mar 01 & NGC~3368 &  $J$ & 15 min & 0.85 & $2\farcs 1 \times 1\farcs 5$ \\
12 Mar 01 & NGC~3368 &  $H$ & 14 min & 0.85 & $2\farcs 4 \times 1\farcs 8$ \\
16 Mar 01 & NGC~3384 & $J$ & 12 min & 0.85 & $2\farcs 5 \times 1\farcs 85$ \\
16 Mar 01 & NGC~3384 &  $H$ & 12 min & 0.85 & $2\farcs 2 \times 1\farcs 8 $ \\
16 Mar 01 & NGC~3384 & $K'$ & 12 min & 0.85 & $2\farcs 1 \times 1\farcs 65$ \\
\hline
\end{tabular}
% \end{center}
\end{flushleft}
\end{table}

We have also observed the global kinematics of the ionized gas in
NGC~3368 with the scanning Fabry-Perot Interferometer (IFP) at the 6m
telescope. In contrast to the integral-field 2D spectrographs MPFS
and SAURON, the IFP allows us to obtain spectral information over a
large field of view, but over a relatively small spectral range.
We use the IFP in interference order 235 (for
$\lambda6563$\,\AA).The IFP is installed at the pupil plane of a focal
reducer attached to the f/4 prime focus of the 6m telescope. A
brief description of this device is available from
http://www.sao.ru/$\sim$gafan/devices/ifp/ifp.htm. The detector was
a CCD TK $1024 \times 1024$ working with a binning of $2\times 2$
pixels. The resulting pixel size was $0\farcs 68$ and the field of view
was about of $5\farcm 8$.
During every object exposure we obtain 32 frames with
interference rings for varying IFP gaps. The full spectral
range (interfringe) was $28$\,\AA, the spectral resolution was
about $2.5$\,\AA. A narrow-band filter with
$FWHM\approx15$\,\AA\ was used to select a spectral domain around
the redshifted galactic emission lines H$\alpha$ and
[NII]$\lambda6583\AA$. The log of the observations with the IFP
is given in Table~3.

Besides the 2D spectral data, we have obtained NIR photometry for
two of the galaxies under consideration. The observations were made at the
2.1m telescope of the National Astronomical Observatory of Mexico
``San Pedro M\'artir'' with the infrared camera CAMILA. The camera is
equipped with a NICMOS3 detector with a format of $256 \times 256$
pixels; mostly the mode with a scale of $0\farcs 85$ (f/4.5) has been used,
except for NGC~3368 which has also been observed with a higher sampling.
The details of the photometric observations are given in Table~4.
Additionally, for all three galaxies we have retrieved the NICMOS/HST
data from the HST Archive. NGC~3368  was
observed on May 4, 1998, with the NIC2 camera, through the filters
F110W and F160W during 128 sec in the framework of a program
of Massimo Stiavelli (ID 7331), and on May 8, 1998, with the NIC2 camera
through the F160W filter during 320 sec in the framework of a program
of John Mulchaey (ID 7330). NGC~3379 was
observed on June 14, 1998, with the NIC3 camera through the F160W
filter during 192 sec  as part of a program of William Sparks
(ID 7919). NGC~3384 was
observed on April 3, 1998, with the NIC2 camera through the F160W
filter during 128 sec for the program of John Tonry (ID 7453).

Almost all the data, spectral and photometric, except the data obtained with
the MPFS, have been reduced with software produced by V.V. Vlasyuk
at the Special Astrophysical Observatory \citep{vlas}. Primary reduction
of the data obtained with the MPFS was done in IDL with software
created by one of us (V.L.A.). The Lick indices were calculated with
our own FORTRAN program as well as by using a FORTRAN program written by
A. Vazdekis.
For the reduction of the IFP data we used our IDL software
\citep{mois_ifp}; also the ADHOC software developed at the Marseille
Observatory by J. Boulesteix (see
http://www-obs.cnrs-mrs.fr/ADHOC/adhoc.html) was used.
The observational data were converted into a ``data cube'' of 32
images. The data reduction includes subtraction of the night
sky emission, channel-by-channel correction \citep{mois_ifp},
wavelength  calibration, and spatial and spectral gaussian smoothing.
The velocity fields and monochromatic images in both emisison
lines (H$\alpha$ and [NII]$\lambda$6583) have been constructed by
Gauss fitting spectral line profiles; also, images in the red
continuum close to the emission lines were calculated from the
same IFP ``data cubes''.

\section{Lick indices and stellar population properties}

Up to now there have been a few attempts to map 2D distributions of
Lick indices by means of integral-field spectroscopy; we
mention here papers by \citet{ems96} on NGC~4594,  by \citet{pel2dfis}
on NGC~3379, 4594, and 4472, and by \citet{n4365sau}
on NGC~4365. In addition we mention our own
papers\citep{sil99b,we2000,we2002,
afsil02a,afsil02b,s0_3} on NGC~7331, NGC~7217, NGC~4429, NGC~7013,
NGC~5055, NGC~4138, NGC~4550, NGC~5574, and NGC~7457.
A difficulty entering reliable mapping of Lick indices comes from the
fact that
we calibrate extended (panoramic) data but are using point-like calibration
sources -- Lick standard stars that are usually placed in the center
of the field-of-view or frame. If the spectral resolution, or the spectral
response, or both, vary over the frame, this may result in a systematic
distortion of the Lick index distributions over the field of view.
We check this effect  by measuring Lick-index distributions
in twilight exposures: the surface index distributions of the H$\beta$,
Mgb, Fe5270, and Fe5335 calculated from twilight frames must be flat, and
the mean level of every index distribution must be close to the values
measured by us earlier: H$\beta _{\odot}=1.86\,$\AA, Mgb$_{\odot}=2.59\,$\AA,
Fe5270$_{\odot}=2.04\,$\AA, and Fe5335$_{\odot}=1.59\,$\AA\ \citep{sharina}.
Additionally, we compare azimuthally-averaged
index radial profiles with well-calibrated long-slit measurements
found in the literature. Figures~2 and 3 present such comparisons for
NGC~3379 and NGC~3384, respectively.

The long-slit data along the major axis of NGC~3379 are taken from
\citet{vazd2}. Their errors exceed the errors of the
azimuthally-averaged SAURON data by several times; however, the
Mgb-profile (Fig.~2, middle) allows us to conclude that there is no
systematic shift between the SAURON- and Vazdekis et al.'s long-slit data
for this particular index. It means that we have properly taken into account,
or have justly neglected, the effects which are not related to the
wavelength (spectral localization) of the features, i.e., the effects of
the spectral resolution and stellar velocity dispersion broadening
or contamination by one of the neighboring spectra  which does not exceed
2\%\--3\%\ according to our estimates. Meantime, the SAURON measurements
of H$\beta$ are marginally too high (by less than 0.2~\AA), and the
SAURON measurements of Fe5270 are certainly too low by 0.3--0.4~\AA.

The radial profiles of the Lick indices in NGC~3384 (Fig.~3) have been
measured even more thoroughly than those in NGC~3379: we have plotted
the azimuthally-averaged MPFS and SAURON data as well as
major- and minor-axis
long-slit measurements from \citet{fish96}. Again, the radial
profile of Mgb looks the most reliable among all the indices --
probably, because the magnesium line falls in the middle of the spectral
range observed. The H$\beta$ measurements outside of the nucleus
according to SAURON are higher
by 0.3--0.4~\AA\  as compared to the MPFS, and in this
particular case the Fisher et al.'s measurements support
the SAURON results. But as for Fe5270, the SAURON data are again
too low by $\sim$0.4~\AA, as seen already in the case of NGC~3379.
\citet{sau2} present preliminary SAURON results
on NGC~3384 and noted this disagreement between their and \citet{fish96}
measurements, but they insisted that their results were
more correct. Since our MPFS data for Fe5270 in NGC~3384 agrees
with that of \citet{fish96} and since the same systematic
shift of Fe5270 is observed in NGC~3379, we suppose that the SAURON
values of the iron index are systematically underestimated.

\begin{table}
\caption[ ]{Linear fits of the Lick index gradients in the bulges}
% \begin{center}
\begin{flushleft}
\begin{tabular}{|l|cc|}
\hline
Lick index & $a$, \AA\ per arcsec & $b$, \AA\ \\
\hline
\multicolumn{3}{|l|}{NGC 3379}\\
Mgb & $-0.0261 \pm 0.0023$ & $4.59 \pm 0.002$ \\
Fe5270 & 0 & $2.64 \pm 0.01$ \\
\hline
\multicolumn{3}{|l|}{NGC 3384}\\
Mgb & $-0.0381 \pm 0.0036$ & $4.04 \pm 0.03$ \\
Fe5270 & $-0.0142 \pm 0.0036$ & $2.61 \pm 0.03$ \\
\hline
\end{tabular}
% \end{center}
\end{flushleft}
\end{table}

However, intrinsically the SAURON azimuthally-averaged data are very
precise: a typical error for every point in the profiles of Figs.~2 and
3 is $0.02 - 0.04$~\AA\ (the accuracy of the MPFS azimuthally-averaged
indices is about 0.1~\AA). Their high quality allows us to diagnose
chemically distinct cores in both galaxies NGC~3379 and NGC~3384:
although the magnesium- and iron-index breaks are of rather low amplitude,
we note a certain change of the profile slopes at $R\approx 4\arcsec$.
Let us also note that we predicted the existence of a chemically distinct
nucleus in NGC~3379 from the multi-aperture photometric data \citep{sil94}.
If we approximate the `bulge' data at $R \ge 4\arcsec$ by linear
fits (the parameters of these fits are given in Table~5), we can obtain
extrapolated `bulge' indices at $R=0\arcsec$, or in the very centers
of NGC~3379 and NGC~3384, and compare them to the real nuclear indices;
the differences would characterize a chemical distinctness of the nuclei.
In NGC~3384 (Fig.~3) $\Delta$Mgb=0.64~\AA\ and $\Delta$Fe5270=0.42~\AA,
$\pm 0.08$~\AA. If we treat the differences of
metal-line indices between the nucleus and the extrapolated bulge as a
difference of metallicity and apply the evolutionary synthesis models for
old stellar populations of \citet{worth94}, we obtain
$\Delta$[Fe/H]=+0.3,
the nucleus being on average more metal-rich, the
$\Delta$Mgb and $\Delta$Fe5270 results being consistent. In NGC~3379
(Fig.~2) the chemical
distinctness of the core is more modest: $\Delta$Mgb=$0.26\pm 0.06$~\AA\
corresponds to $\Delta$[m/H]=+0.1, and $\Delta$Fe5270=0.08~\AA\
implies an even smaller $\Delta$[Fe/H]$\approx +0.05$, which is evidence for
a marginally higher magnesium-to-iron ratio in the`nucleus' as
compared to the `bulge'.

High-precision radial profiles of absorption-line indices provide
a zero-order diagnostic of the central stellar substructures;
but with 2D index distributions we are able to discuss a morphology
of the chemically distinct entities and their relation to photometric
substructures, in particular to compact circumnuclear disks. Let us
consider in detail every galaxy of our small sample.

\subsection{NGC 3379}

Figure~4 presents Lick-index maps for the central part of NGC~3379.
As expected, the chemically distinct nucleus is better seen
in Mgb; the central maximum of Fe5270 is of rather low contrast, and
the distribution of H$\beta$ looks almost flat. The core distinguished
by the enhanced magnesium line is well-resolved and has an elongated
shape. After heavy smoothing (Fig.~4, upper right) we are able
to determine an orientation of this elongated magnesium-rich core:
$PA_{core}\approx 84^{\circ}$. The magnesium-index isolines trace
a substructure which is more flat than the main stellar body: the axis ratio
of the magnesium-index isolines is $b/a_{Mg} \approx 0.5$, whereas the
isophote axis ratio is $b/a_{phot}\approx 0.9$. Is it a compact,
$R \le 4\arcsec$, circumnuclear disk that is seen as a chemically
distinct core?

Figure~5 allows us to quantify characteristics of the stellar populations
in the center of NGC~3379 at different distances from the center.
The magnesium-iron diagram (Fig.~5, top)
reveals a systematic shift of the galactic measurements with respect
to a model locus of the populations with the solar magnesium-to-iron
ratio plotted according to \citet{worth94}. Although we know already
from the analysis of Fig.~2 that the SAURON data of Fe5270 are slightly
too low with respect to the standard Lick system, the systematic
shift of the NGC~3379 measurements relative to Worthey's models
is so large that it cannot be explained by any index system bias.
For example, the data of \citet{trager} for the very center
of NGC~3379 and the sequence obtained by substitution of the \citet{vazd2}
long-slit data on Fe5270 in the radial profile table
(Fig.~5, top) are also away from the model locus. We have to conclude that
the nucleus (core) of NGC~3379 is magnesium overabundant,
[Mg/Fe]$=+0.2 - +0.3$, and this enhanced Mg/Fe ratio holds to an approximately
constant level (within 0.05 dex) as a function of radius.

So to determine an age of the stellar population in the center of
NGC~3379, we must use the models with [Mg/Fe]$>0$; such as the models
of \citet{tantalo} that are calculated in particular for
[Mg/Fe]$=+0.3$. But the models of \citet{tantalo} involve the
combined iron index $\langle \mbox{Fe} \rangle \equiv$(Fe5270+Fe5335)/2,
and we have only Fe5270 which is unfortunately biased. By plotting
an age-diagnostic diagram (Fig.~5, bottom), we have confronted
the $\langle \mbox{Fe} \rangle$ measurements of \citet{vazd2},
which are well-calibrated on to the Lick system, though are not very
precise, to our measurements of H$\beta$ at the same distances from the
center, which on the contrary have a small error.
Since the accuracy of the age estimates is determined
mainly by the H$\beta$ accuracy, we obtain a certain mean
(luminosity-weighted) age estimate of 8--9~Gyr for the very center
($R< 8\arcsec$) of NGC~3379, with possible variations within one Gyr
with radius. The mean age of the stellar population in the
core of NGC~3379 below 10~Gyr is in agreement with the recent result
of \citet{terlfor} who have found an integrated value
of 9.3~Gyr within an aperture of $R_e/8$.

\subsection{NGC 3384}

Figure~6 presents Lick index maps for the center of NGC~3384
which have been obtained with the MPFS, and Fig.~7 presents
similar maps obtained with SAURON for a larger area; note that
the iron-index map covers the full area whereas the preliminary results of
\citet{sau2} showed only half of the field of view
because of some problems they had with the data reduction. The
spatial resolution of the former map is better by a factor of two than
that of the latter, so in Fig.~6 we see a lot of subtle
details which are not seen in Fig.~7. However, common features
can also be noted: the magnesium index demonstrates a central
maximum, rather compact and at high-contrast, and well-resolved.
Also, the iron enhancement in the center of NGC~3384 is more diffuse
and extended, so even from the analysis of Figs.~6 and 7 ``by eye''
it is suggestive that there is a gradient of the magnesium-to-iron ratio
with radius. The metal-line enhanced areas are elongated as if
they were produced by a compact circumnuclear disk appearance; however
when we drew isolines of the Mgb distribution (Fig.~7, upper right),
we convinced ourselves that their orientation, $PA_{Mg} \approx
20^{\circ} - 25^{\circ}$, differs substantially from the orientation
of the inner isophotes, $PA_{inner}\approx 40^{\circ} - 45^{\circ}$
(see the next Section), and from the line of nodes, $PA_0=53^{\circ}$
(Table~1). The H$\beta$ distribution demonstrates an unresolved peak
in the nucleus in the MPFS data only (Fig.~6, left), whereas
the SAURON data is evidence for quite a flat distribution (Fig.~7, lower
right).

The diagnostic diagrams for the average data at different radii are
shown in Fig.~8, which includes only the MPFS data as they offer
unbiased iron-index estimates. These diagrams present a slightly different
picture from that in the center of NGC~3379. Indeed, there exists
a Mg/Fe ratio gradient (Fig.~8, top): the nucleus is obviously
magnesium-overabundant, $\mbox{[Mg/Fe]}_{nuc} \approx +0.3$ if one takes
into account its very young mean age (see Fig.~8, bottom); and at
$R \ge 5\arcsec$ the Mg/Fe ratio is about solar. Consequently, to
determine a mean age, we must use two sets of stellar population models,
with [Mg/Fe]$=+0.3$ and with [Mg/Fe]=0. Figure~8, bottom, presents
the comparison of our data to the models of \citet{tantalo}
for both values of [Mg/Fe]. One can see that the unresolved nucleus
of NGC~3384 is rather young for a lenticular host: its mean stellar age
is less than 5~Gyr, and probably close to 3~Gyr; the general metallicity
is higher than the solar one: $\mbox{[m/H]}_{nuc} \approx +0.3 - +0.4$.
In the nearest vicinity of the nucleus the metallicity drops to the
quasi-solar value, and the mean age increases to 7--8~Gyr. Therefore,
although in the center of NGC~3384 an extended region looks chemically
distinct (the net dimensions of this region are difficult to determine
because they are comparable to our resolution limit), the unresolved
stellar nucleus has probably followed its own, quite separate evolution.

\subsection{NGC 3368}

The maps of the metal-line Lick indices for the central part of NGC~3368
(Fig.~9) demonstrate a qualitative difference with respect to the maps
of the centers of NGC~3379 and NGC~3384: instead of peaks in the nuclei
they show `holes' -- net minima of Mgb and
$\langle \mbox{Fe} \rangle$ just where the surface brightness has a
maximum. These holes may be evidences for a metal deficiency of the
nuclear stellar population, or more likely, for its extreme youth.
Magnesium- and iron-index enhancements can also be seen in the maps
of Fig.~9, especially on the February MPFS maps (Fig.9, top) which have
a better spatial resolution, but these peaks are off-nuclear. The Mgb
index has two maxima at the major axis located at $R\approx 4\arcsec$
symmetrically with respect to the nucleus, and the iron index
$\langle \mbox{Fe} \rangle$ demonstrates something like a half-ring
with a radius of $2\arcsec - 3\arcsec$. Such features associated
usually to star-forming rings and `bright circumnuclear spots' are clear
signatures of a bar presence; indeed, the isophotal twist in the center
of NGC~3368 can be noticed even over the very limited area covered by
the MPFS frame. In the recent study of NGC~3368 structure and
kinematics by \citet{mois02} the presence of the minibar with
an extension of about $5\arcsec$ has also been noted.

The age diagnostics in the center of NGC~3368 is complicated because of
rather intense ionized-gas emission: the H$\beta$ absorption-line
index is contaminated by the Balmer emission line. We have tried to
take into account the effect of H$\beta$ in emission when calculating
azimuthally-averaged index profiles: we have co-added separately the blue
and the red spectra in the same concentric rings; then, by using the red
spectra at various radii, we estimated equivalent widths of the H$\alpha$
emission line which is stronger than H$\beta$ and can be surely measured
even inside a deep H$\alpha$ absorption line; after that we
calculated the H$\beta$-index correction by involving the mean relation
between H$\alpha$ and H$\beta$ emission lines in normal galaxies,
$EW(H\beta em)=0.25EW(H\alpha em)$ \citep{sts2001}. The
corrected indices taken as a function of radius are plotted in the diagnostic
diagrams presented in Fig.~10.
The correction applied to the H$\beta$ index is close to the minimal
possible one (corresponding to pure radiative ionization), so the
derived ages may be slightly overestimated. By considering both
diagrams in Fig.~10 together, one can conclude that the nuclear stellar
population has a mean age of 3~Gyr and a mild magnesium overabundance,
[Mg/Fe]$\approx +0.2$. When moving toward $R\approx 10\arcsec$ in radius,
the mean age strongly increases, and the magnesium-to-iron ratio
drops to the solar value.

The ages that we determine by confronting the measured integrated-spectra
indices to simple (one-age, one-metallicity) stellar
population models are indeed mean luminosity-weighted ages: if one has a mix
of populations of different ages, each of them would contribute to the
integrated spectrum proportionally to its luminosity, and from the
diagnostic diagrams like that in Fig.~10 (bottom), one could
obtain an age estimate, intermediate between extreme individual stellar
generation ages. So although we have obtained a mean age of 3~Gyr
for the nuclear stellar population in NGC~3368, it does not mean that
all stars there have been formed 3~Gyr ago; it means that we see
a superposition of an intermediate-age stellar generation onto the older
stellar bulge. To illustrate this idea, we have constructed a model
experiment for the age-diagnostic diagram of Fig.~10: to a moderately
metal-poor ([Fe/H]=-0.22), old ($T=12$ Gyr) bulge we added a
relatively young ``poststarburst'' population with [Fe/H]$=+0.25$ and
two trial ages -- 1 and 3~Gyr, with a contribution from 1\%\ to
80\%\ of the total mass. The calculations were made with
the `Dial-a-model' machine of Guy Worthey available at his WEB-page
(http://astro.wsu.edu/worthey/). Two
model sequences corresponding to $T_{burst}=1$ and 3~Gyr were plotted
in Fig.~10 (bottom), along which the contribution of the young population
varies. One can see that although the former sequence crosses
the SSP-model line for $T_{SSP}=3$~Gyr at a value
of $\sim 20$\%\, the overall trend of the $T=1$~Gyr young starburst
deviates from the observational points, whereas the
$T_{burst}=3$~Gyr sequence coincides exactly with the radial trend in
the index in the center of NGC~3368. So we conclude that the age of the
secondary nuclear star formation burst of 1~Gyr or less can be excluded
and that whereas $T_{nuc}=3$~Gyr is the mean luminosity-weighted
estimate, the true age of the secondary starburst does not differ
strongly from this value.

\section{Stellar and gaseous kinematics in the centers of the
          Leo I Group galaxies}

Since integral-field spectroscopy provides us with two-dimensional
line-of-sight velocity fields, we are able now to analyse both
the rotation and central structure of the galaxies.
If we have an axisymmetric mass distribution
and rotation on circular orbits, the direction of maximum central
line-of-sight velocity gradient (we shall call it the ``kinematical major
axis'') should coincide with the line of nodes as well as the
photometric major axis, whereas in a case of a triaxial
potential the isovelocities align with the principal axis of the ellipsoid
and the kinematical and photometric major axes generally diverge,
showing twists in an opposite sense with respect to the line of nodes
\citep{mbe92,mm2000}.
In a simple case of cylindric (disk-like) rotation we have a convenient
analytical expression for the azimuthal dependence of central
line-of-sight velocity gradients within the area of solid-body rotation:\\

\noindent
$dv_r/dr = \omega$ sin $i$ cos $(PA - PA_0)$, \\

\noindent
where $\omega$ is the deprojected central angular rotation velocity,
$i$ is the inclination of the plane, and $PA_0$ is the
orientation of the line of nodes, coinciding in the case of an
axisymmetric ellipsoid (or a thin disk) with the photometric
major axis. So by fitting azimuthal variations
of the central line-of-sight velocity gradients with a cosine
curve, we can determine the orientation of the kinematical major
axis by its phase and the central angular rotation velocity
by its amplitude. This is our main tool for kinematical analysis.

Let us note that the method of the analysis of line-of-sight velocity
gradients works correctly only within the area of solid-body rotation.
Therefore
it cannot be used for a large-scale velocity field (beyond the central
kpc region). For analysis of the IFP's velocity fields we applied a method
usually referred to as the `tilted-ring' method \citep{begeman}. In the
framework of this
method the velocities  were fitted in elliptical rings (elongated along the
PA of the
galactic disk major axis) by a model of pure circular rotation.
As a first step we found and fixed the rotation center position which is
the center of symmetry of the velocity fields. As a second step we
calculated the radial dependence of the model parameters: systemic velocity
$V_{sys}$, rotational velocity $V_{rot}$, disk inclination $i$ and position
angle of the kinematical major axis $PA_0$. And finally,  $i$ and
$V_{sys}$ were fixed at their mean values and the run of $V_{rot}(r)$ and
$PA_0(r)$ with radius were obtained. Radial variations of
$PA_0$, if present, can be used for detecting various types of non-circular
motions -- see \citet{mm2000} for details and references.

\subsection{NGC 3379}

NGC~3379 was treated earlier as an example of a classic elliptical
galaxy -- round and homogeneous. However, recent kinematic studies
by using a long-slit technique \citep{n3379ls,it_3379}
have proved that at least the central part of the galaxy
consists of several subsystems including a highly inclined dust (and
gaseous?) ring and that NGC~3379 may be a misclassified S0. Since we
analyse the results of 2D kinematic mapping, new details showed up
to complicate the picture.

Figure~11 shows the stellar line-of-sight velocity field and the stellar
velocity dispersion field measured from the SAURON data. The former field
demonstrates a rather fast (for an elliptical galaxy), quasi-axisymmetric
rotation; the latter field reveals a prominent maximum in the center with
a quasi-axisymmetric radial decrease of the velocity dispersion --both
facts were
already noted, e.g. by \citet{n3379ls}. We have measured
an orientation of the kinematical major axis up to $R\approx 5\arcsec$,
which is the approximate edge of the rapidly rotating area, and have found
that it changes significantly from a $PA_{kyn}=265^{\circ}$ ($85^{\circ}$)
at $R\approx 2\arcsec$ to $PA_{kyn}=256^{\circ}$ ($76^{\circ}$) at
$R=4\arcsec - 5\arcsec$. We compare the orientation of the kinematical
major axis to that of the photometric major axis according to the data
of \citet{frei} and to the NICMOS/HST image analysis results
in Fig.~12; we would like also to refer to the results of the analysis
of the WFPC2/HST images of NGC 3379 by \citet{it_3379}.
Within $R\le 15\arcsec$ the photometric major axis deviates
in a positive sense from the tabulated outer isophote orientation
$PA_0=70^{\circ}$ (see the Table~1) and agrees well with the kinematical
major axis. Let us also remind here that the isolines of the enhanced
magnesium
index deviate from the outer isophote orientation too, although they are
aligned with the
kinematical major axis; and they have a more flattened distribution
than the isophotes.
Such coincidences are evidence for a compact stellar disk in the center of
NGC~3379 which may be inclined to the main symmetry plane of the galaxy.
There exists another inclined circumnuclear disk in NGC~3379 -- a dust-gaseous
one, with a radius of $2\arcsec$ and $PA_0=125^{\circ}$ \citep{it_3379}.
Are they related? In our Fig.~12, and also in Fig.~3 of \citet{it_3379},
based on the high-resolution results of the HST/WFPC2 F555W and F814W
isophote analysis,
one can see a strong twist of the major axis when approaching
the center, inside the central arcsecond. \citet{it_3379} thought it to be
an effect of the dust ring projection. But the dust
ring with an orientation of $PA=125^{\circ}$ would force an isophote twist
in a polar direction, so in a negative sense with respect to the line
of nodes $PA_0 =70^{\circ}$. Such a turn is really observed in the radius
range of $0\farcs 8 - 1\farcs 5$ (Fig.~12). However within $R=0\farcs 4$
the effect of the dust ring having a radius of $2\arcsec$ is negligible.
We would think the circumnuclear isophote twist of $PA\approx 102^{\circ}$
in our measurements, or even up to $PA\approx 120^{\circ}$, as the rectified
analysis of \citet{it_3379} suggests, to be real
and to represent a signature of the inclined circumnuclear stellar disk
which is probably related to the dust-gaseous one. Some discrepancies of the
position angle estimates can be explained by different spatial
resolutions of the data.

\subsection{NGC 3384}

Figure~13 presents the stellar line-of-sight velocity field and that
of the stellar velocity dispersion in the center of NGC~3384, according
to the data from MPFS (top) and SAURON (bottom). The stellar LOS-velocity
field reveals a fast regular rotation, with isovelocity features typical
for a disk embedded into a massive bulge (tight isovelocity crowding
near the line of nodes, etc). The presence of the compact circumnuclear
stellar disk with a radius of $6\arcsec$ in the center of NGC~3384
was found earlier from a photometric analysis by \citet{it2_3384},
so we confirm their finding from a kinematical point of view. Yet another
exotic subcomponent -- a stellar polar ring suggested by \citet{fr_3384}
and \citet{polar_r} -- can be discarded with certainly based on the
same kinematical arguments. This latter feature is localized in the
radius range of $10\arcsec - 25\arcsec$, and the large SAURON velocity
map for the stars in the center of NGC~3384 does not show any switch
of the spin orientation (toward the polar direction) at $R=10\arcsec -
15\arcsec$. Hence we diagnose this morphological feature (identified as
``EC'' in the
terminology of Busarello et al.) as a bar. Bars belong always
to disks (cold dynamical subsystems), so we conclude that the inclination
of $90^{\circ}$ given for NGC~3384 by LEDA (see the Table~1) is erroneous:
one would not see a bar along the minor axis as an elongated structure
under edge-on orientation of the global disk; in the meantime, the refined
analysis of \citet{it2_3384} has demonstrated a very elongated,
almost ``peanut''-shape structure, with $PA=132^{\circ}$, in this radius
range. We show radial variations of the characteristics of global isophotes
in NGC~3384 obtained through the NIR broad-band filters in Fig.~14. The
sharp maximum of isophote position angle at $R=15\arcsec$ and the
corresponding minimum of the ellipticity refer to a superposition of
a flattened spheroid with $PA_0\approx 50^{\circ}$ and of a bar almost
along the minor axis. An asymptotic ellipticity at $R>60\arcsec$,
$1-b/a=0.45$, implies a possible disk inclination as low as $57^{\circ}$,
(\citet{it1_3384} obtained $i=63.5^{\circ}$). If the symmetry plane
of NGC~3384 is inclined by some $60^{\circ}$ to the line of sight, the
proposed ``polar ring'' is indeed a bar in the disk plane with a radial
extension of $\sim 2.5$ kpc ($40\arcsec$). The presence of the bar in
NGC~3384 is confirmed by the stellar velocity dispersion distribution
in the center of the galaxy (Fig.~13): it has a strongly elongated shape
with a flat maximum aligned roughly with the minor axis. As \citet{vd97}
have shown from dynamical modelling, within the bar potential
the stellar velocity dispersion distributions are elongated along the bars.
Just this picture is observed in the center of NGC~3384.

Let us quantify the kinematical arguments in favor of the circumnuclear
disk. In Fig.~15 we compare orientations of the photometric (according
to the HST data) and the kinematical (according to the MPFS and the SAURON
data) major axes within $R\approx 5\arcsec$ from the center. The agreement
is good within $1^{\circ} - 2^{\circ}$, or certainly to within our accuracy
limits. However both photometric and kinematic major axes
($PA_{phot}=46^{\circ}$ ($226^{\circ}$) and
$PA_{kyn}=225.5^{\circ} \pm 1.2^{\circ}$) deviate significantly
(and consistently!) from the line of nodes, $PA_0=53^{\circ}$.
Such behavior is evidence for an {\em inclined} circumnuclear stellar disk.
Interestingly, as the orientation of the bar is $PA_{bar}=132^{\circ}$
\citep{it2_3384}, obviously we deal with a disk {\em which is polar
with respect
to the bar}. Such a configuration can be often encountered when
studying circumnuclear
gas rotation within a triaxial potential \citep{we99,we2000}.
Here for the first time we have faced
a similar, but completely stellar substructure.

\subsection{NGC 3368}

The kinematical picture in the center of NGC~3368 can be presented in more
detail than that in NGC~3379 and NGC~3384 because its intense emission
lines provide us with the velocity field of the ionized gas. However more
information does not clarify the situation but rather makes it quite puzzling.

Figure~16 presents the stellar velocity field and the stellar velocity
dispersion distribution in the center of NGC 3368 according to the MPFS
data of February 2000; the analogous data of the March 2000 observations
are described by \citet{mois02}. The isovelocities demonstrate
a regular solid-body rotation of stars; but the kinematical major axis
is strongly different from the orientation of the inner isophotes
which have a noticeable twist of the major axis over the radius range of
$0\arcsec -8\arcsec$, unlike the isovelocities. The stellar velocity
dispersion map is rich with subtle details and looks quite unusual:
the nucleus is not distinguished, neither by a peak nor by a minimum
of $\sigma _{*}$, but two arc-like regions -- one of low stellar velocity
dispersion at $R\approx 4\arcsec - 6\arcsec$ and another of high
stellar velocity dispersion at a larger radius  -- prevent any
reasonable interpretation of this data.

Figure~17 shows the ionized-gas velocity field and the 2D distributions of
the emission-line surface
brightness for [NII]$\lambda$6583 and H$\alpha$ which are obtained with MPFS.
Again, the gas rotates rather regularly and as a solid body up to the edges
of the area investigated. As for the emission maps, here there is a surprise:
whereas the [NII] emission distribution repeats roughly the shape of the
continuum
isophotes, the H$\alpha$ emission-line intensity distribution is elongated
orthogonally, approximately along the minor axis of the inner isophotes. Is
this structure real? Let us appeal to high-resolution imagery. Figure~18
shows, firstly, the WFPC2/HST map of NGC~3368 in the visual regime and
secondly,
the NICMOS/HST color map (close to $J-H$). One can see a quasi-polar compact
dust ring encircling the north-west part of the central elongated structure
(of a minibar?). The dust (and gas?) density in this polar ring is so high
that it is seen very well even at wavelengths longer than 1~mkm; it manifests
itself at the color map as a red lane aligned in $PA\approx 30^{\circ} -
35^{\circ}$ to the north-west from the blue nucleus. Since this dust polar
ring is traced by the H$\alpha$ emission, but is unseen in the [NII] emission
line, we may suppose that it is a site of intense current star formation. Why
is it polar? Inner gas polar structures are often met in the galaxies with a
triaxial central potential as we have already mentioned when considering
NGC~3384.

Let us quantify, as is usually done, a characteristic of rotation of the
stars and ionized
gas in the center of NGC~3368. In Fig.~19 we have plotted some selected
azimuthal dependencies of the central LOS velocity gradients for the stars
in radius ranges of $2\farcs 0 - 2\farcs 8$ and of $3\farcs
0 - 4\farcs 2$ and
for the ionized gas in a radius range of $2\farcs 3 - 3\farcs 3$; for the
stars we have plotted both the February and the March MPFS data, their
agreement being excellent. The stars rotate slightly slower than the gas:
$19\pm 2$  km/s/arcsec versus $26.4 \pm 4.2$ km/s/arcsec. The long-slit
observations by \citet{vega} also confirm this  kinematical
feature. It is natural because the stellar velocity dispersion in the center
of NGC~3368 is larger than 100 km/s, and the emission lines are narrow.
However the orientation of the
kinematical major axes for the ionized gas and for the stars are rather
similar: $176^{\circ} \pm 2^{\circ}$ for the former and $167.5^{\circ} \pm
1.5^{\circ}$ for the latter subsystem. Indeed, we have no serious reasons to
treat the gas and stellar rotations as different -- they may rotate together.
The coincidence of the kinematical major axes near
$PA_{kyn} \approx 170^{\circ}$, an orientation which is not marked in
any way in the central part of NGC~3368, is very strange. One could
understand a rotation of the central ionized gas
in the plane towards $PA=170^{\circ}$, because the outermost neutral
hydrogen is distributed in an extended disk with just this line of nodes
\citep{leoh1_3}. Meanwhile the molecular gas in the central part of the galaxy
demonstrates elliptical isodensities elongated in $PA\approx 40^{\circ}$
\citep{co99} -- the direction that coincides with the line of nodes
of the dust polar ring; but the velocity field of the molecular gas over
the entire area covered by mapping resembles rather a prolate rotation, with
$PA_{kyn} \approx 170^{\circ}$ \citep{co99}. So, once more, for the
ionized gas the orientation of the rotational plane at
$PA_{kyn}=170^{\circ} -175^{\circ}$ is kinematically understandable; but
among the visible {\it stellar} structures in NGC~3368 there are none
with a distinguishing orientation at this position angle -- see Figs.~20
and 21.
Figure~20 shows a comparison of the NICMOS/HST isophote characteristics
with orientation of the kinematical major axes. The
$PA_{kyn}\approx 170^{\circ}$ holds in the radius range of $2\arcsec -
5\arcsec$ whereas the orientation of the isophotes holds
at $PA_{phot}\approx 125^{\circ}$ (closer to the nucleus the isophotes
twist by more than $100^{\circ}$, and we think this twist to be caused by
an effect of the dust polar ring projection). \citet{mois02}
have argued from the azimuthal Fourier analysis of the large-scale surface
brightness distributions that the line of nodes of the inner ($R<140\arcsec$)
disk of NGC~3368 is $PA_0=135^{\circ}$ and that the value of $5^{\circ}$
given by LEDA may be related only to the very outermost part of the galaxy.
So we are convinced that in the center of NGC~3368 we see a minibar with
a semimajor axis of $5\arcsec$; however non-circular streaming motions
around the bar with $PA_{bar}=125^{\circ}$ located in the disk with
$PA_0=135^{\circ}$ would cause a twist of the kinematical major axis
from the line of nodes in a positive sense, as observed, but only
by about $10^{\circ}$ \citep{mbe92}. Meanwhile it turns by at least
$30^{\circ}$, and therefore this turn cannot be explained by a
simple dynamical effect of minibar triaxiality.
Do we observe a rotation of a rather young faint stellar subsystem
which is formed from the circumnuclear gas and shares its spin?
It would be possible if this subsystem is much colder than
the bulge dominating the surface brightness.

Fig.~21 shows large-scale radial variations of the isophote
characteristics of NGC~3368 to illustrate its very complex structure.
While the surface brightness profile of the galaxy looks very regular,
with a single-scale exponential disk dominating the bulge at
$R > 22\arcsec$ \citep{kent85}, the isophote orientation and ellipticity
change all the way up to the optical edges of the galaxy, implying
either disk oval distortions or a strong disk warp.
As mentioned in Section~2 we used the scanning Fabry-Perot data to
study the large-scale ($r=100\arcsec -200\arcsec$) kinematics of the
ionized gas in
two emission lines (H$\alpha$ and [NII]$\lambda$6583). The velocity fields
in both lines are shown in Fig.~22. Unfortunately stellar absorption
and interference orders which overlap distort the emission line profiles
in the H$\alpha$ datacube over the central region of the galaxy. We cannot
resolve this problem because the free spectral range of the IFP is small
($\sim28$\,\AA). Therefore the H$\alpha$ line-of-sight velocities were
measured only in the star-forming ring (r=$50\arcsec -90\arcsec$) and in
fragments of spiral arms. But the [NII] velocity field was constructed over
the full range in radius (Fig.~22). The shape and orientation of the
isovelocity
contours at $r<7\arcsec$ demonstrate an agreement with the MPFS velocity
field in [NII]$\lambda$6583 of Fig~17.
The `tilted-ring' analysis of our Fabry-Perot velocity fields
confirms the stable orientation of the ionized gas kinematical major axis at
$PA_{kyn}=170^{\circ} -175^{\circ}$ over the global disk of the galaxy
out to a distance of $200\arcsec$ from the center. So the global
dust-gaseous disk of the galaxy is probably decoupled completely from the
stellar one, having its own orientation in space. Since an HI bridge is
seen between the intergalactic gaseous ring and NGC~3368 \citep{leoh1_2},
we suggest that all the gas in this spiral galaxy is of recent
external origin and is not yet relaxed with respect to the main symmetry plane
of the galaxy. We must stress the good agreement
between the measurements of the $PA_{kyn}$ in two independent IFP velocity
fields; but also we note a discrepancy of the rotational velocity amplitudes.
The differences between the rotational velocities in the H$\alpha$ and
[NII] (Fig.22) may be explained in the frame of the hypothesis of
the inclined gaseous disk. As it is shown in Fig.22, the [NII] rotation
velocity at $r=50\arcsec -90\arcsec$ is systematically slower by
$\sim 50\, \mbox{km} \cdot \mbox{s}^{-1}$ than the H$\alpha$ one
(under the assumption
of a disk inclination of $i=48^\circ$). If the gaseous disk is inclined
to the main galaxy plane then shock-wave fronts can exist at the cross-section
of the global stellar and gaseous disks, because gas strikes
the gravitational well.  Moreover, the strong stellar spiral arms in this
region intensify the contrast of the gravitational potential. Therefore the
low
[NII] velocities may be explained if the part of the nitrogen emission line
is emitted by shock-excited gas slowed down by collisions with the stellar
disk and spiral arms.

\section{Discussion}

We have considered the three brightest galaxies in the vicinity of the
intergalactic gaseous ring in the Leo~I group. We have found that all three
galaxies have peculiarities, or extra components, in their centers; and the
space orientations of these extra components are related to the line
of nodes of the intergalactic gaseous ring, $PA_{ring}\approx 40^{\circ}$,
in all three cases.
NGC~3384 has a circumnuclear stellar disk with the
line of nodes at $PA=45^{\circ}$, inclined to the main symmetry plane
of the galaxy but aligned with the global intergalactic gaseous ring.
NGC~3368 has a polar (relative to the central minibar)
dust-gaseous ring with a projected major axis at $PA\approx 35^{\circ}$,
and its CO distribution is also elongated in $PA\approx 40^{\circ}$.
NGC~3379 has a circumnuclear dust-(and gaseous) ring (and probably a stellar
disk) with the major axis at $PA=120^{\circ} - 125^{\circ}$, orthogonal
to the intergalactic ring orientation; but it is the only ellitical galaxy
among ours, with a clear signature of triaxiality \citep{n3379ls},
so its triaxial main body may provoke gas transfer on to a polar orbit
when spiraling to the center.
These are too many coincidences, if one takes into account the diversity
of the
global major axis orientations (see Table~1). We can suggest that
all the circumnuclear structures mentioned above have the origin
related to accretion of the gas from the intergalactic ring; in this
case the spin conservation would provide alignment of the circumnuclear
disks at $PA_0=40^{\circ}$, except in NGC~3379.
Geometry and sizes of the circumnuclear structures
must depend strongly on the global structures of the galaxies under
consideration, in particular on their triaxiality and spheroid/disk ratios.
And finally, the closely related mean ages, 3~Gyr, of the nuclear stellar
populations in NGC~3368 and NGC~3384 (in NGC~3379 it may seem older
because of the larger contribution of the old spheroid to the integrated
spectrum) imply that a characteristic time between accretion events,
or a timescale for interaction between the galaxies and the intergalactic
ring, is of the order of several Gyrs, or of order the revolution time of
the intergalactic ring which is 4~Gyr, or of order of the crossing time
in the Leo~I group. This result rejects conclusively the scenario of
\citet{roodwil} who proposed a collision between NGC~3368
and NGC~3384 only 0.5~Gyr ago.

\acknowledgements
We are very grateful to Dr. Elias Brinks for editing the manuscript.
Also we thank Dr. S.N. Dodonov for supporting the observations of NGC~3384
at the 6m telescope. The 6m telescope is operated with  financial
support from the Science Ministry of Russia (registration number 01-43);
we thank also the Programme Committee of the 6m telescope for allocating
observing time. During the data analysis stage we have
used the Lyon-Meudon Extragalactic Database (LEDA) supplied by the
LEDA team at the CRAL-Observatoire de Lyon (France) and the NASA/IPAC
Extragalactic Database (NED) which is operated by the Jet Propulsion
Laboratory, California Institute of Technology, under contract with
the National Aeronautics and Space Administration. This work
is partially based on data taken from the ING Archive of the
UK Astronomy Data Centre and on observations made with the NASA/ESA
Hubble Space Telescope, obtained from the data archive at the Space
Telescope Science Institute. STScI is operated by the Association of
Universities for Research in Astronomy, Inc., under NASA contract
NAS 5-26555. The study of the young nuclei in lenticular galaxies
is supported by a grant of the Russian Foundation for Basic Research
01-02-16767 and the study of the evolution of galactic centers --
by the Federal Scientific-Technical Program -- contract
of the Science Ministry of Russia no.40.022.1.1.1101.

\clearpage

\begin{figure}
\plotone{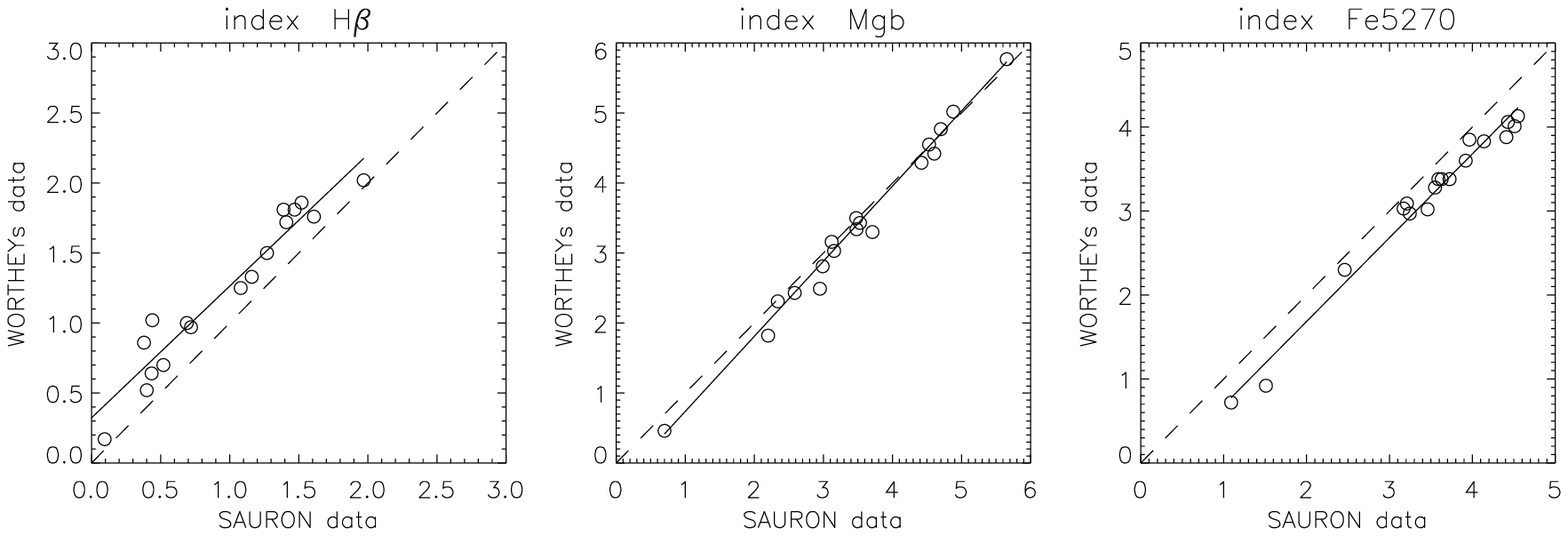}
\caption{The Lick-system index calibration of the SAURON data
obtained during the observing run of March/April 2000;
18 stars from the list of Worthey et al.(1994) are used.
Dashed lines are the bisectors of the quadrants, solid straight lines
are the regression lines calculated by the least-square method}
\end{figure}

\begin{figure}
\epsscale{0.8}
\plotone{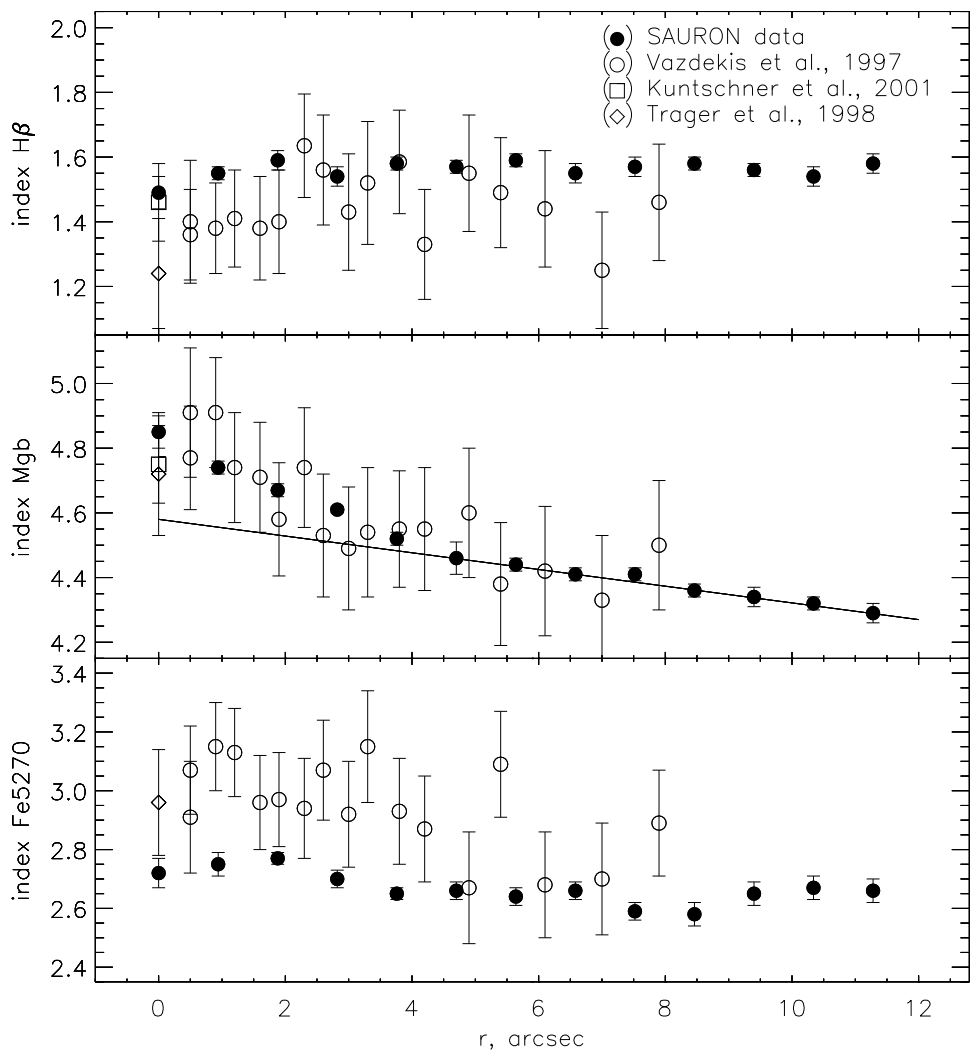}
\caption{Comparison of the SAURON set
of the azimuthally-averaged index measurements for NGC 3379
with the long-slit data taken along the major axis by
Vazdekis et al.(1997). The central $4\arcsec$-aperture
measurements from Trager et al.(1998) and Kuntschner et al.(2001)
are also plotted. To stress the break of the Mgb behavior at $R\approx
5\arcsec$
the straight line fitted by the least-square method to the
SAURON data at $R=5\arcsec - 12\arcsec$ is drawn in the middle plot}
\end{figure}

\begin{figure}
\plotone{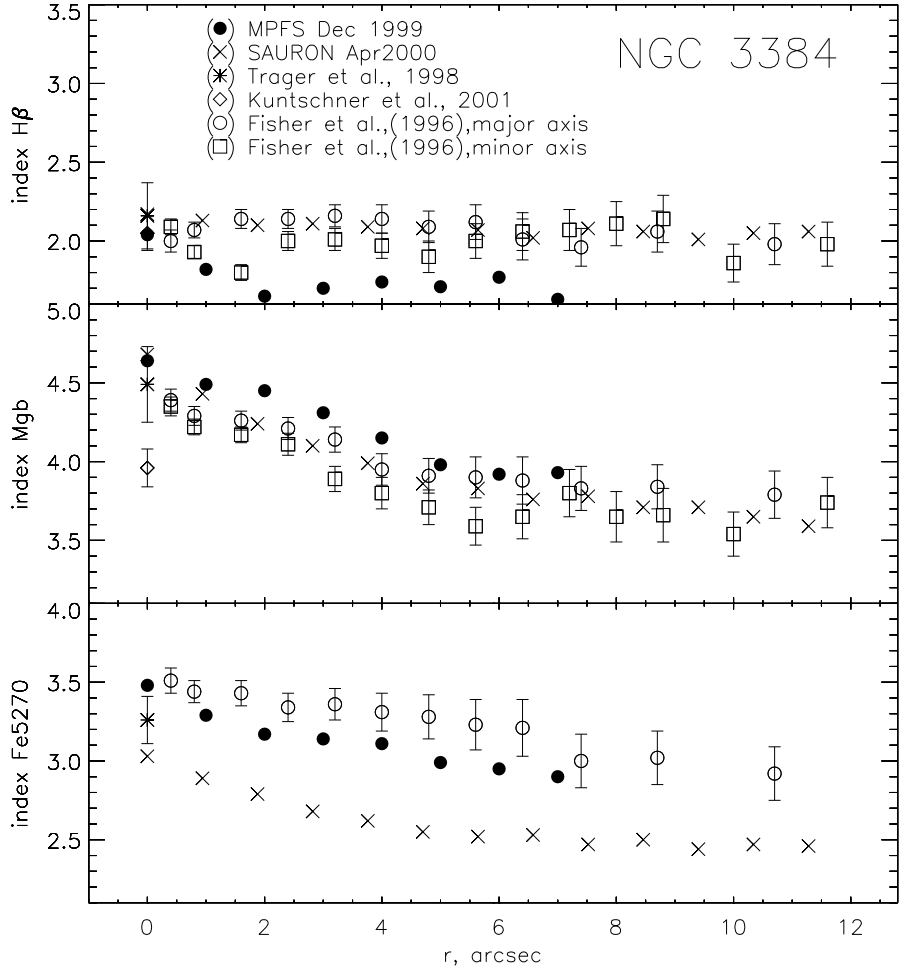}
\caption{The comparison of the SAURON and MPFS sets
of the azimuthally-averaged index measurements for NGC 3384
with the long-slit data taken along the major and minor axes by
Fisher et al.(1996). The central $4\arcsec$-aperture
measurements from Trager et al.(1998) and Kuntschner et al.(2001)
are also plotted}
\end{figure}

\begin{figure}
\plotone{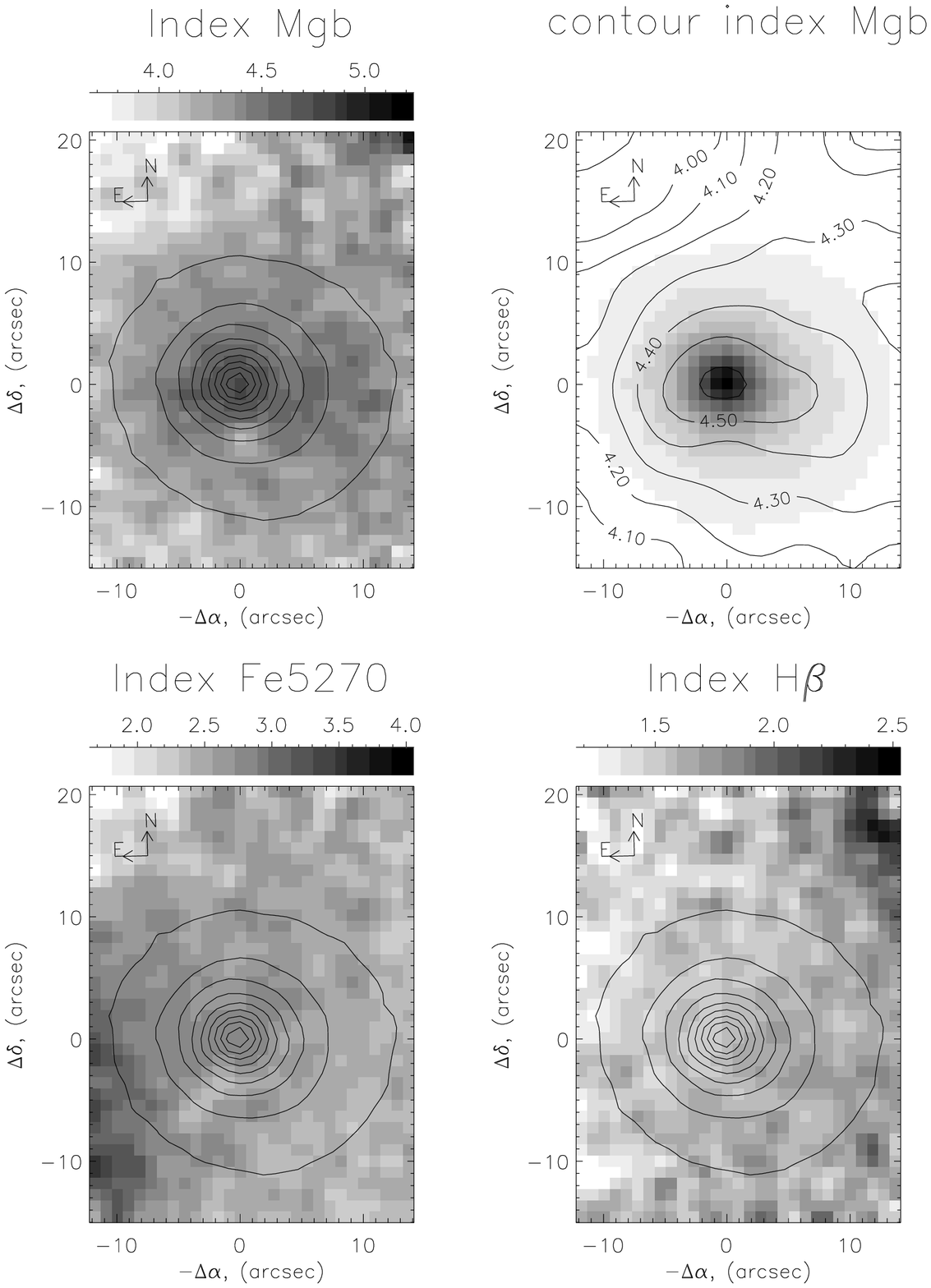}
\caption{The SAURON index maps for NGC 3379. A green
($\lambda$5000~\AA ) continuum is overlaid as isophotes in
three plots out of four. At the top right the Mgb index distribution,
smoothed heavily with a 2D Gaussian of $FWHM=3\farcs 5$ is plotted
by isolines to show the orientation of the chemically decoupled
structure; here the green continuum is gray-scaled. In all plots
the North is up, the East is left}
\end{figure}

\begin{figure}
\epsscale{0.6}
\plotone{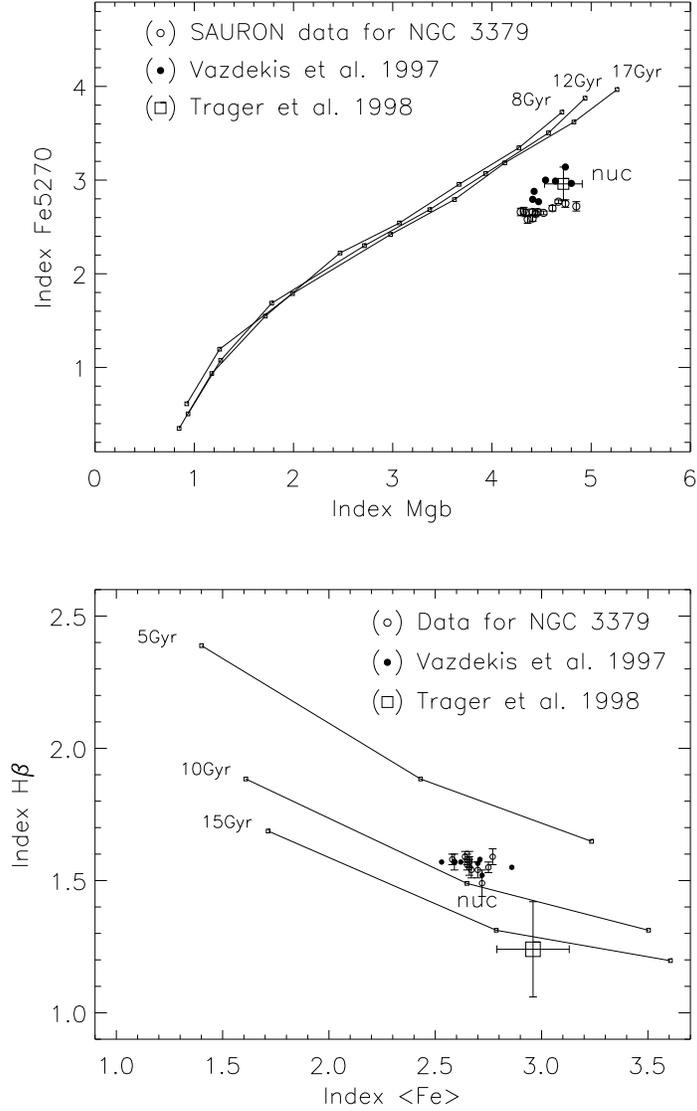}
\caption{The `index vs index' diagnostic diagrams for the
azimuthally-averaged SAURON index measurements and long-slit data
from Vazdekis et al.(1997) for NGC 3379.
The SAURON data points
(open circles) are taken along the radius of the galaxy with a step
of $0\farcs 94$. As a reference frame, the model equal-age sequences
from Worthey (1994) for old stellar populations with [Mg/Fe]=0 are
shown in the top plot as small signs connected by thin lines;
the metallicities for Worthey's models are +0.50, +0.25, 0.00,
--0.22, --0.50, --1.00,--1.50, --2.00, if one takes the signs from
the right to the left. Since the stellar population in NGC 3379,
according to the top plot, has [Mg/Fe]$>0$, the models of Tantalo
et al.(1998) for [Mg/Fe]$=+0.3$ are used for the age diagnostics
at the bottom plot; for these models the metallicities are +0.4, 0.0,
and -0.7}
\end{figure}

\begin{figure}
\epsscale{1}
\plotone{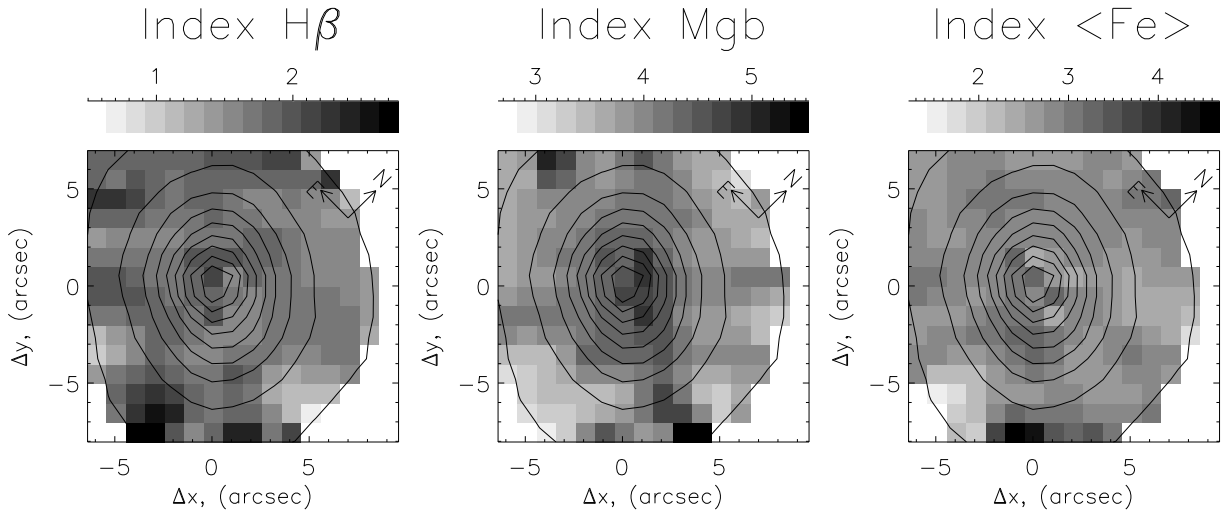}
\caption{The MPFS index maps for NGC 3384
($\langle \mbox{Fe} \rangle \equiv$(Fe5270+Fe5335)/2). A green
($\lambda$5000~\AA ) continuum is overlaid by isophotes.
The images are direct, $PA(top)=46^{\circ}$}
\end{figure}

\begin{figure}
\plotone{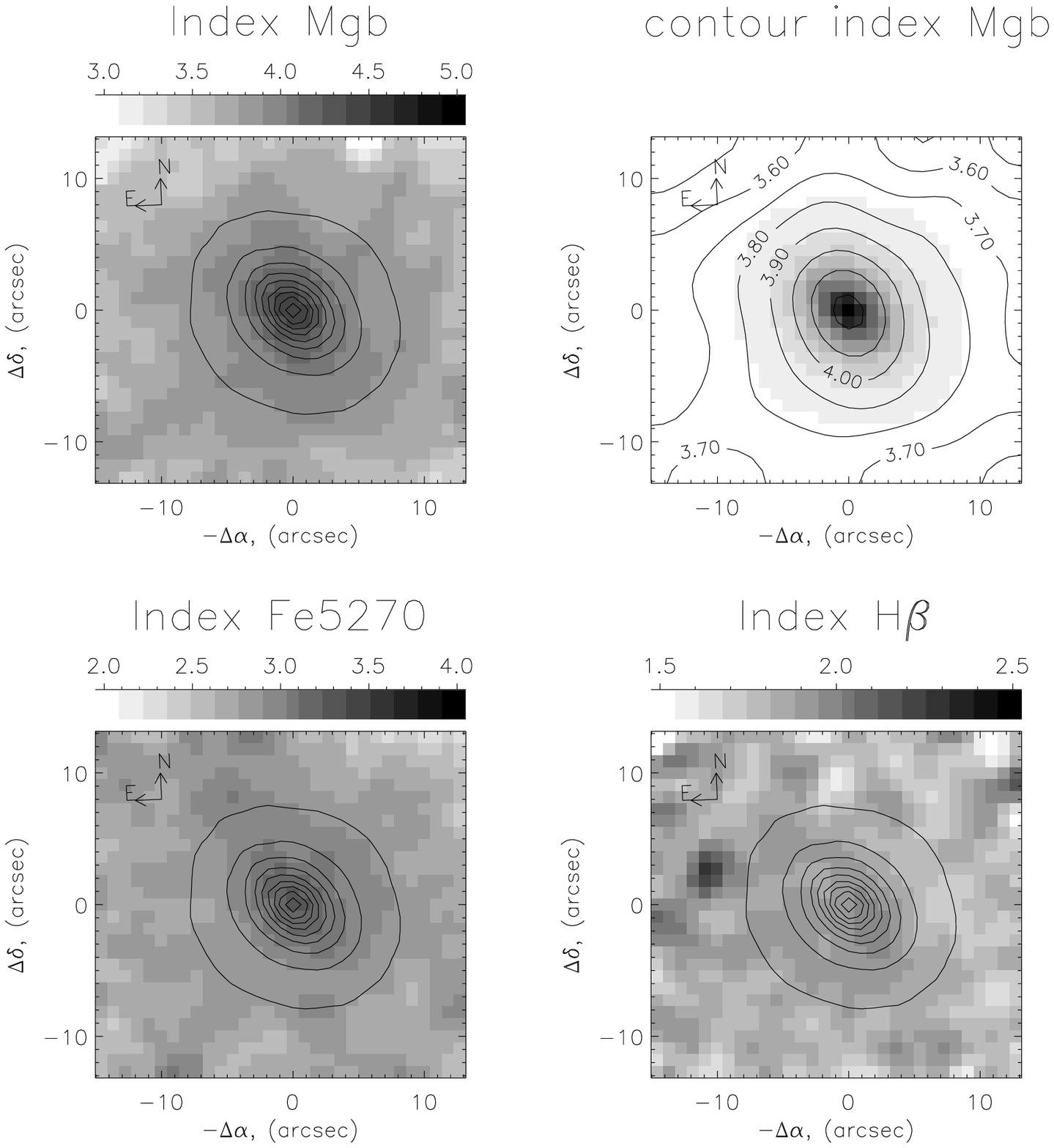}
\caption{The SAURON index maps for NGC 3384. A green
($\lambda$5000~\AA ) continuum image is overlaid as isophotes in
three plots out of four. At the top right the Mgb index distribution
smoothed heavily with a 2D Gaussian of $FWHM=3\farcs 5$ is plotted
by isolines to show the orientation of the chemically decoupled
structure; here the green continuum is gray-scaled. In all plots
the North is up, the East is left}
\end{figure}

\begin{figure}
\epsscale{0.5}
\plotone{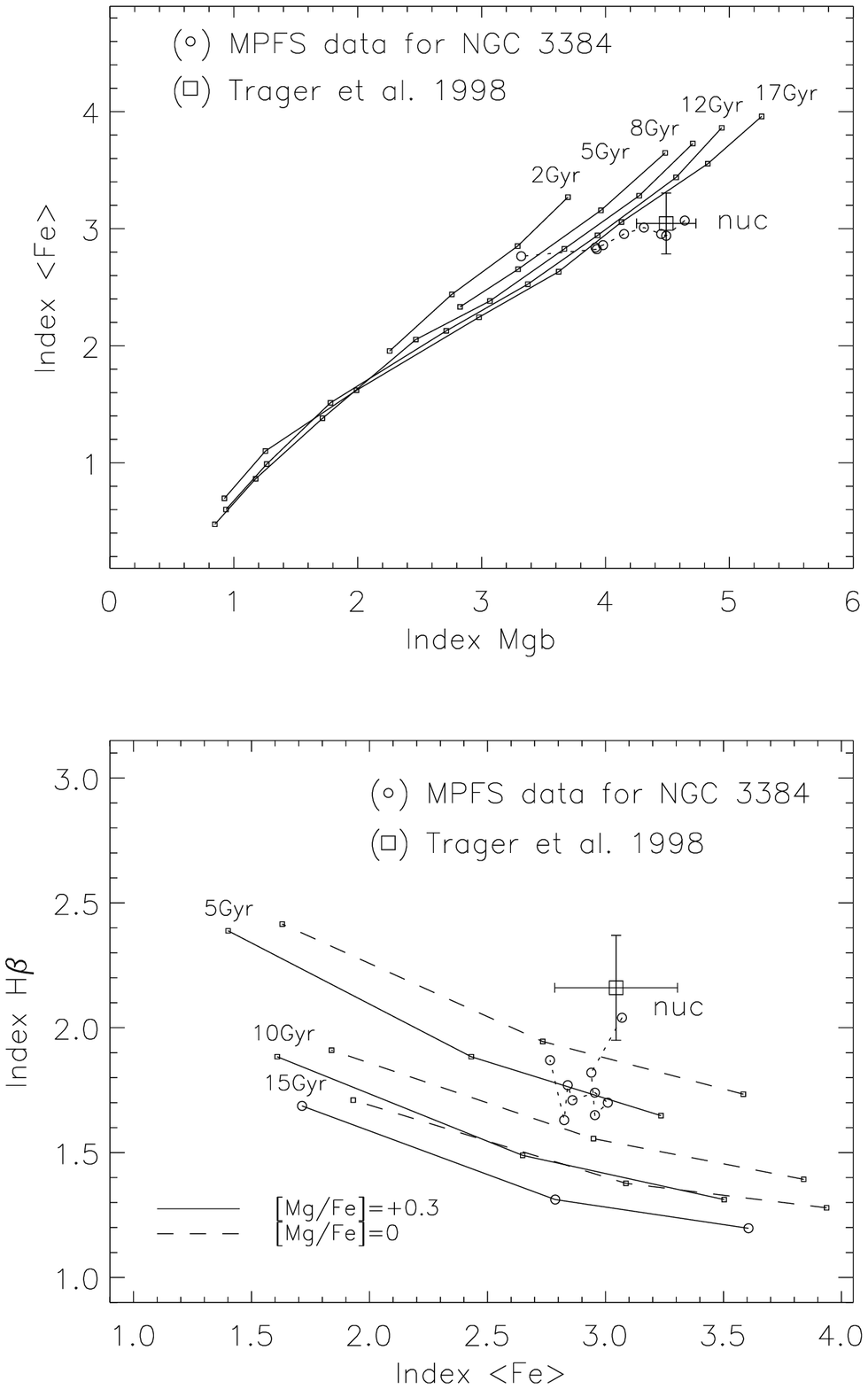}
\caption{The `index vs index' diagnostic diagrams for the
azimuthally-averaged MPFS index measurements in NGC 3384.
The MPFS data points
(open circles) are taken along the radius of the galaxy with a step
of $1\arcsec$. As a reference, the model equal-age sequences
from Worthey (1994) for old stellar populations with [Mg/Fe]=0 are
shown in the top plot as small signs connected via thin lines;
the metallicities for Worthey's models are +0.50, +0.25, 0.00,
--0.22, --0.50, --1.00,--1.50, --2.00, if one takes the signs from
the right to the left (only the four first for the younger models).
Since the stellar population in the nucleus of NGC 3384,
according to the top plot, has [Mg/Fe]$>0$, the models of Tantalo
et al.(1998) for [Mg/Fe]$=+0.3$ are used for the age diagnostics in
the nucleus and the models of the same authors for [Mg/Fe]=0 for
the off-nuclear points in the bottom plot; for these models
the metallicities are +0.4, 0.0, and -0.7}
\end{figure}

\clearpage

\begin{figure}
\epsscale{1}
\plotone{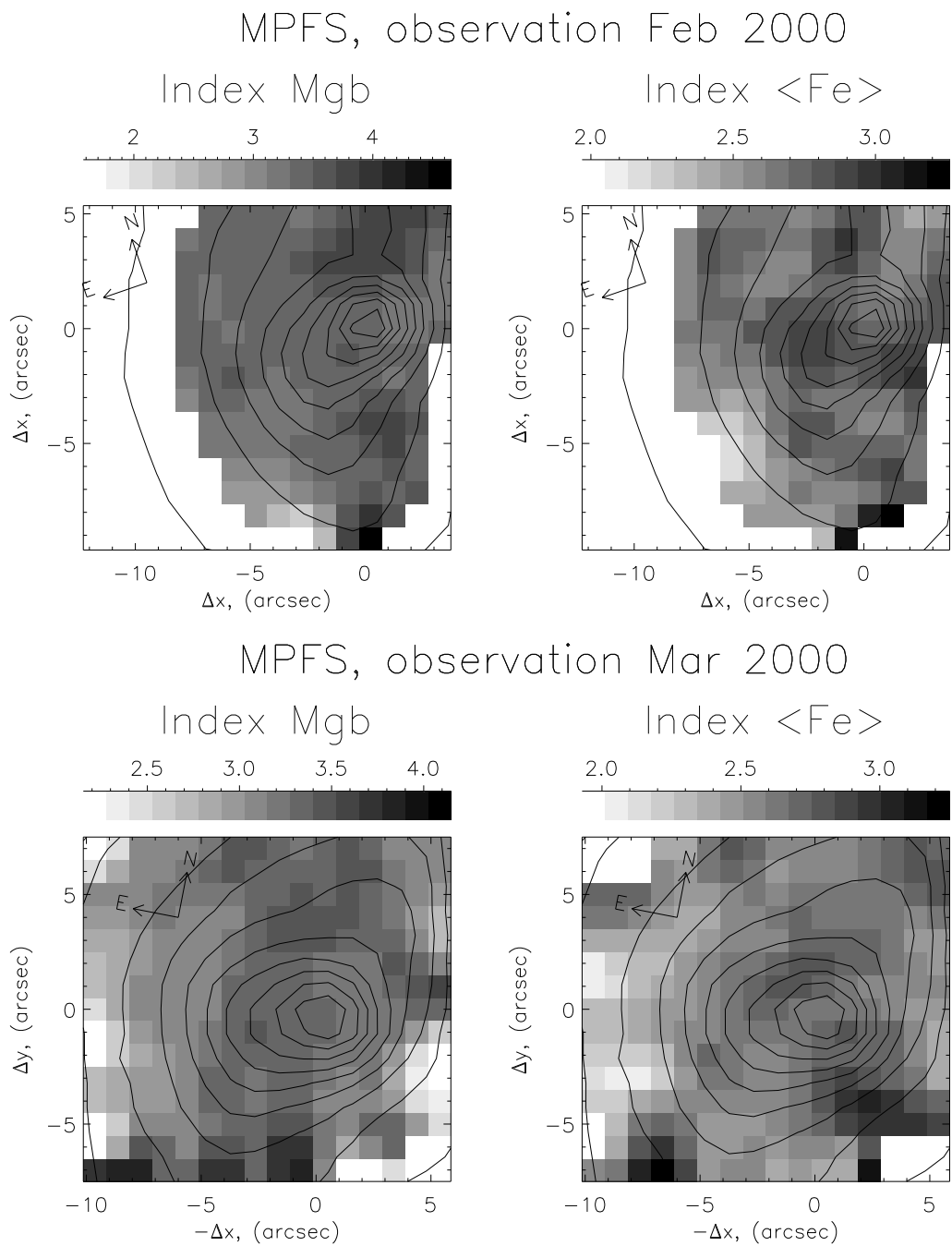}
\caption{The MPFS index maps for NGC 3368: Mgb and
$\langle \mbox{Fe} \rangle (\equiv$(Fe5270+Fe5335)/2) for two
different observing dates. A green
($\lambda$5000~\AA ) continuum is overlaid as isophotes.
The images are direct, in the top row plots $PA(top)=-19^{\circ}$,
in the bottom row plots $PA(top)=+11^{\circ}$}
\end{figure}

\begin{figure}
\epsscale{0.6}
\plotone{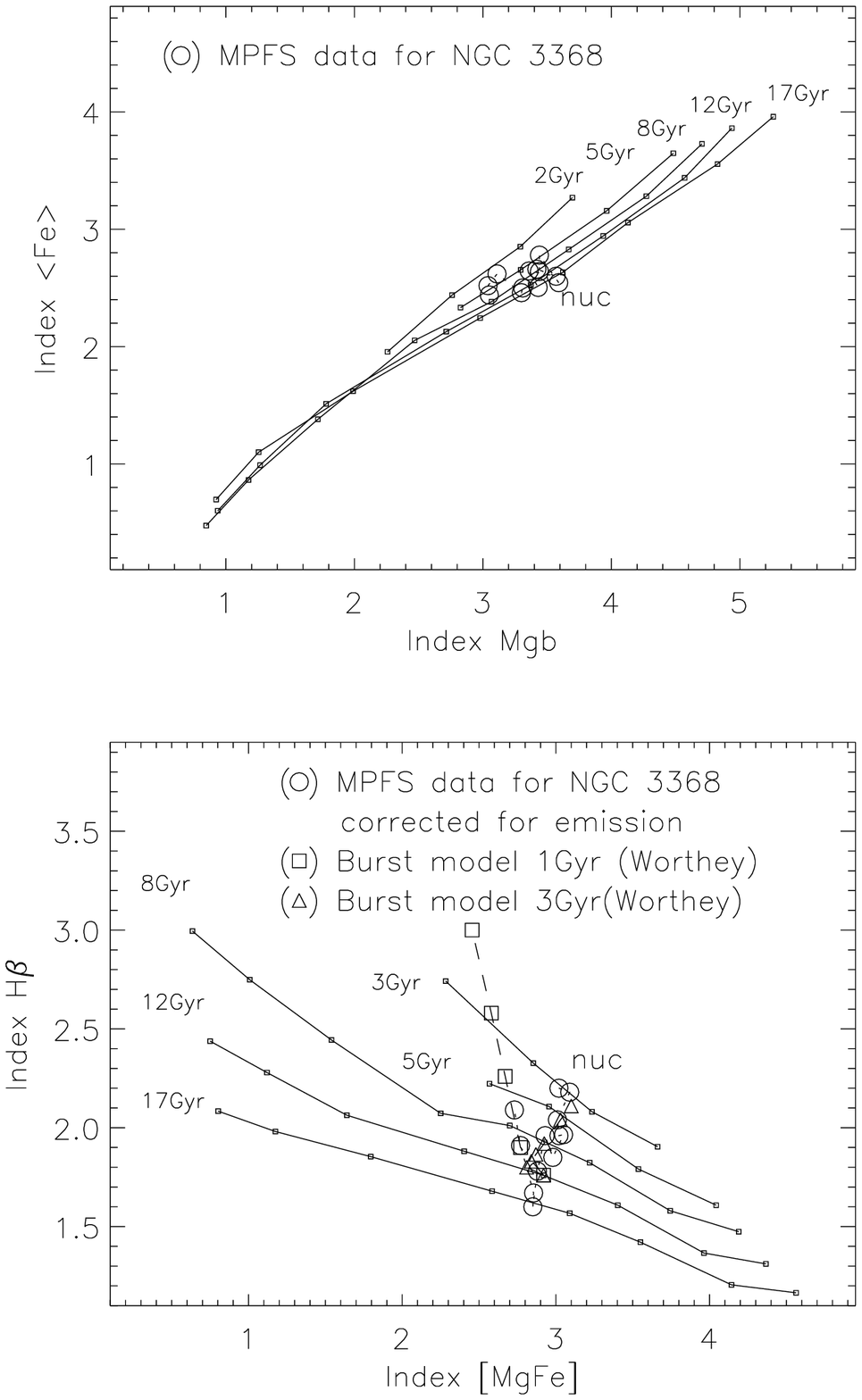}
\caption{The `index vs index' diagnostic diagrams for the
azimuthally-averaged MPFS index measurements in NGC 3368.
The MPFS data points
(open circles) are taken along the radius of the galaxy with a step
of $1\arcsec$. As a reference, the model equal-age sequences
from Worthey (1994) for old stellar populations with [Mg/Fe]=0 are
shown as small signs connected by thin lines;
the metallicities for Worthey's models are +0.50, +0.25, 0.00,
--0.22, --0.50, --1.00,--1.50, --2.00, if one takes the signs from
the right to the left (only the first four for the younger models).
At the bottom plot, which is age-diagnostic diagram, the H$\beta$ index
measurements are corrected for the embedded emission as described in the
text. The "post-starburst models" are also shown, as open squares and
triangles:
the "young population"  of 1 or 3 Gyr old and with [Fe/H]=+0.25 is
superposed onto the "old population" of 12 Gyr and [Fe/H]=-0.22
in various proportion, from 1\%\ to 30\%\ (for a 1-Gyr old burst) or to 80\%\
(for a 3-Gyr old burst)}
\end{figure}

\begin{figure}
\epsscale{1}
\plotone{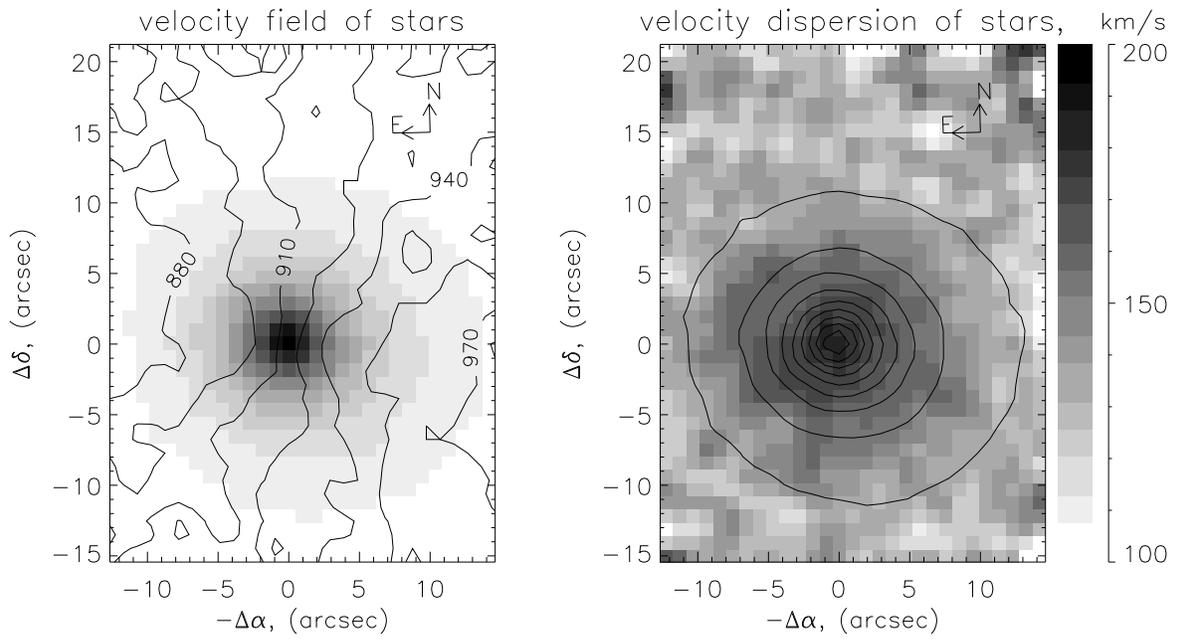}
\caption{Line-of-sight velocity field for the stellar
component (isovelocity lines overlaid onto the continuum image) and stellar
velocity dispersion map (grayscale, with the green continuum isophotes
overlaid) in NGC~3379 according to the SAURON data. In all plots
the North is up, the East is left}
\end{figure}

\begin{figure}
\epsscale{0.8}
\plotone{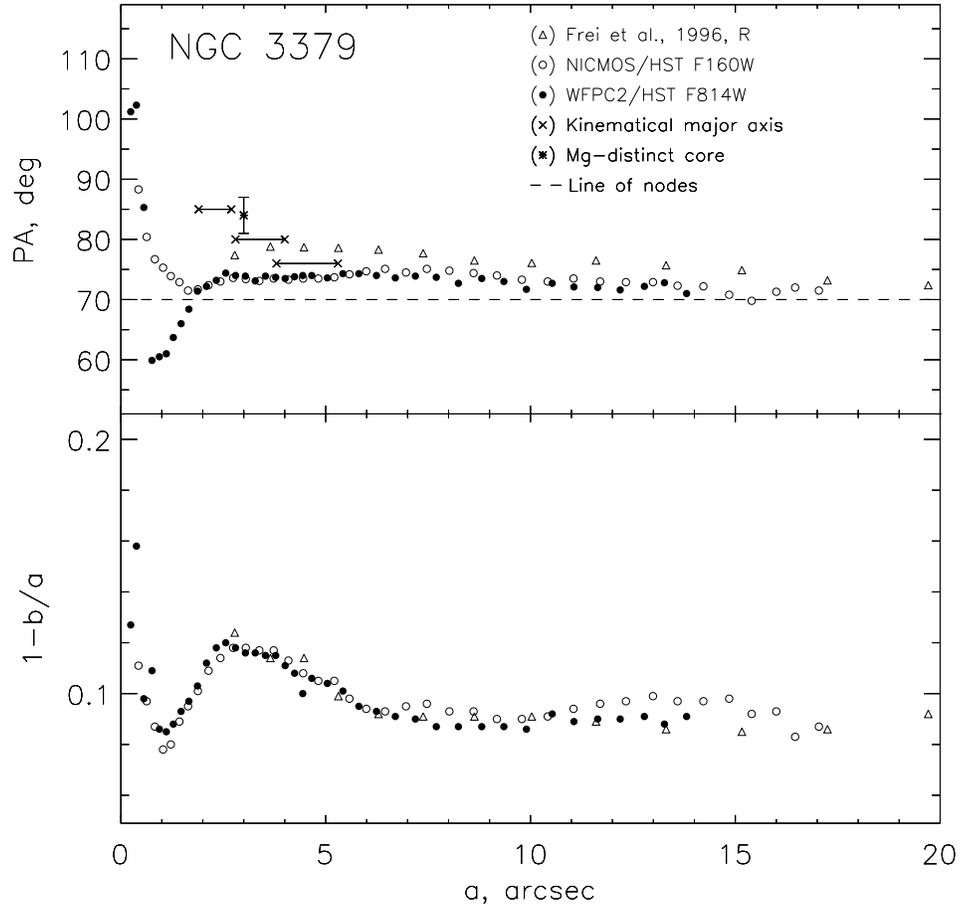}
\caption{Comparison of the orientations of the isophotal
(photometric) major axis and the kinematical major axis (see the text)
in NGC~3379 and radial variations of the isophotal ellipticity}
\end{figure}

\begin{figure}
\epsscale{1}
\plotone{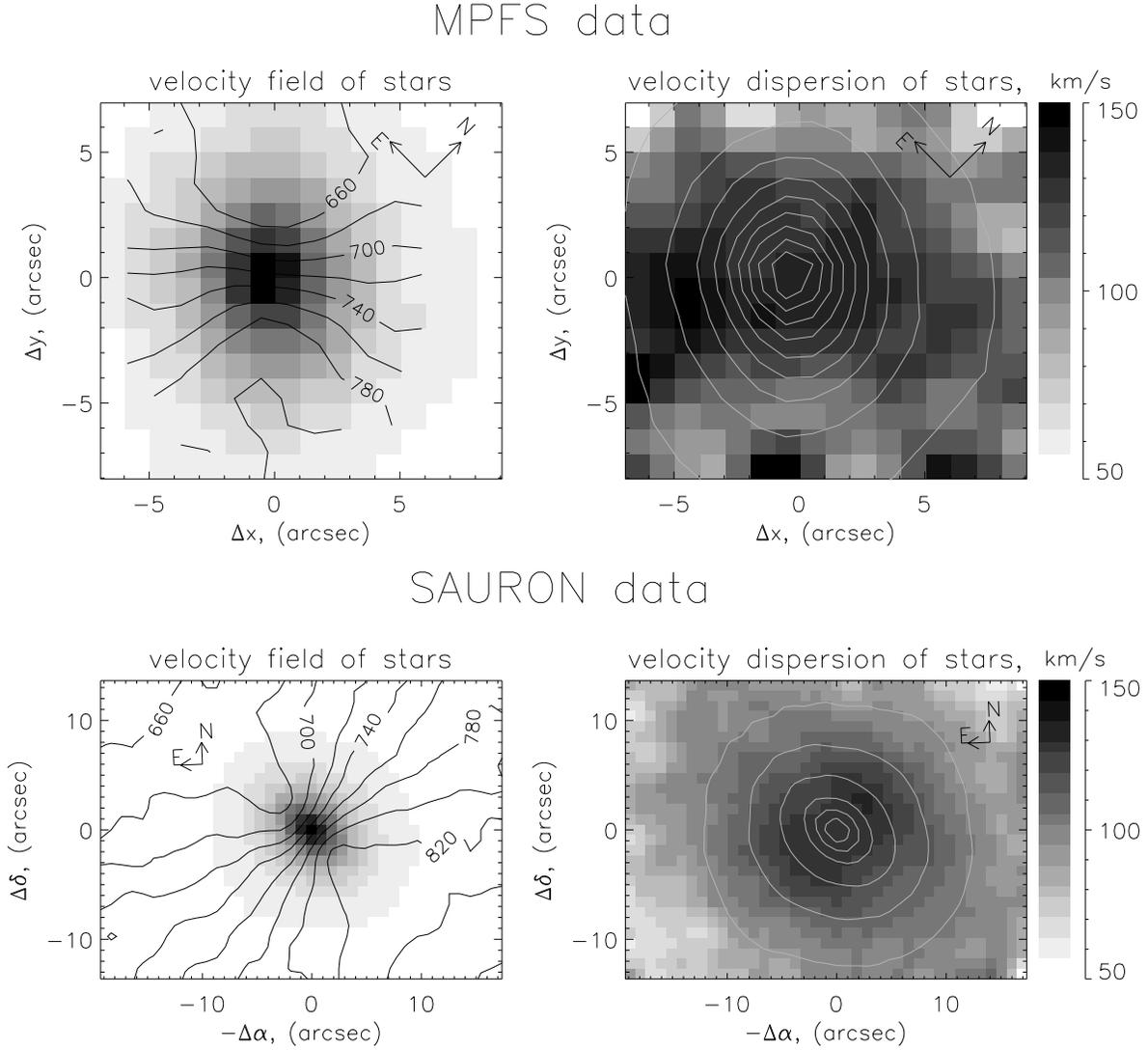}
\caption{Line-of-sight velocity field for the stellar
component (isovelocity lines overlaid onto the continuum image) and stellar
velocity dispersion map (grayscale, with the green continuum isophotes
overlaid) in NGC~3384 according to the MPFS (top) and SAURON (bottom) data.
The images are direct, in the top row plots $PA(top)=+46^{\circ}$,
in the bottom row plots $PA(top)=-2^{\circ}$}
\end{figure}

\begin{figure}
\epsscale{0.8}
\plotone{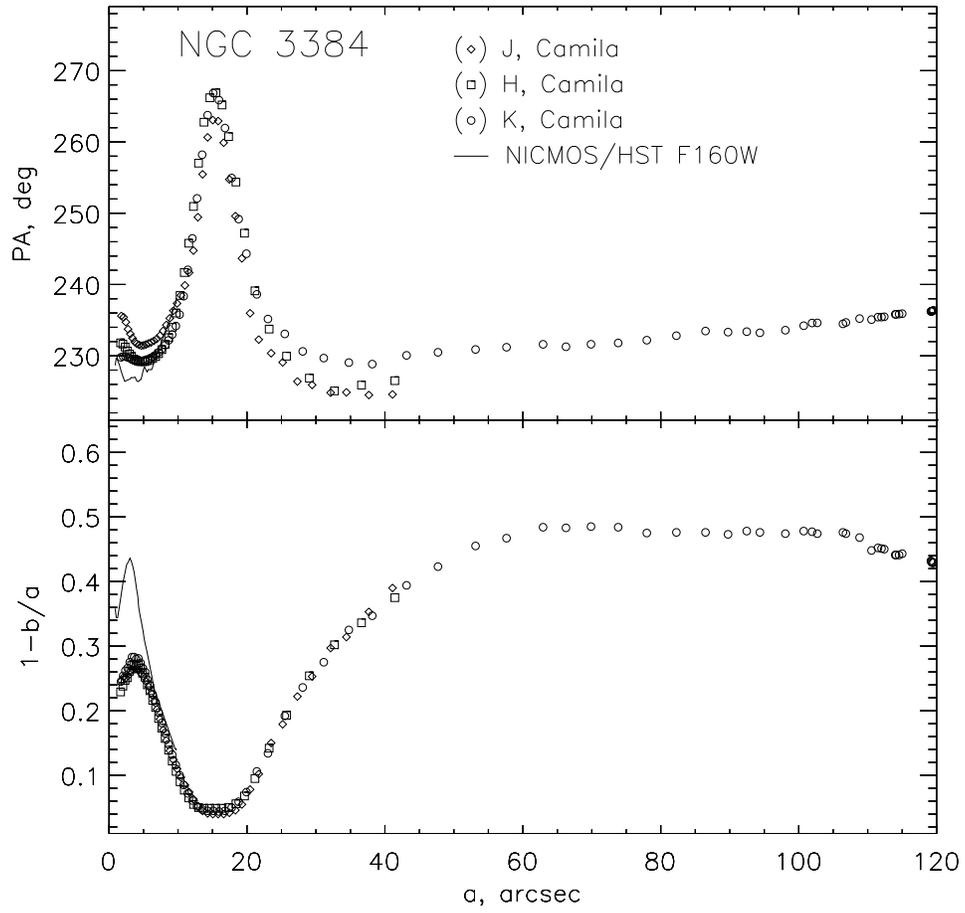}
\caption{Radial variations of the isophotal characteristics in NGC 3384
according to our and to NICMOS/HST near-infrared photometric data}
\end{figure}

\begin{figure}
\plotone{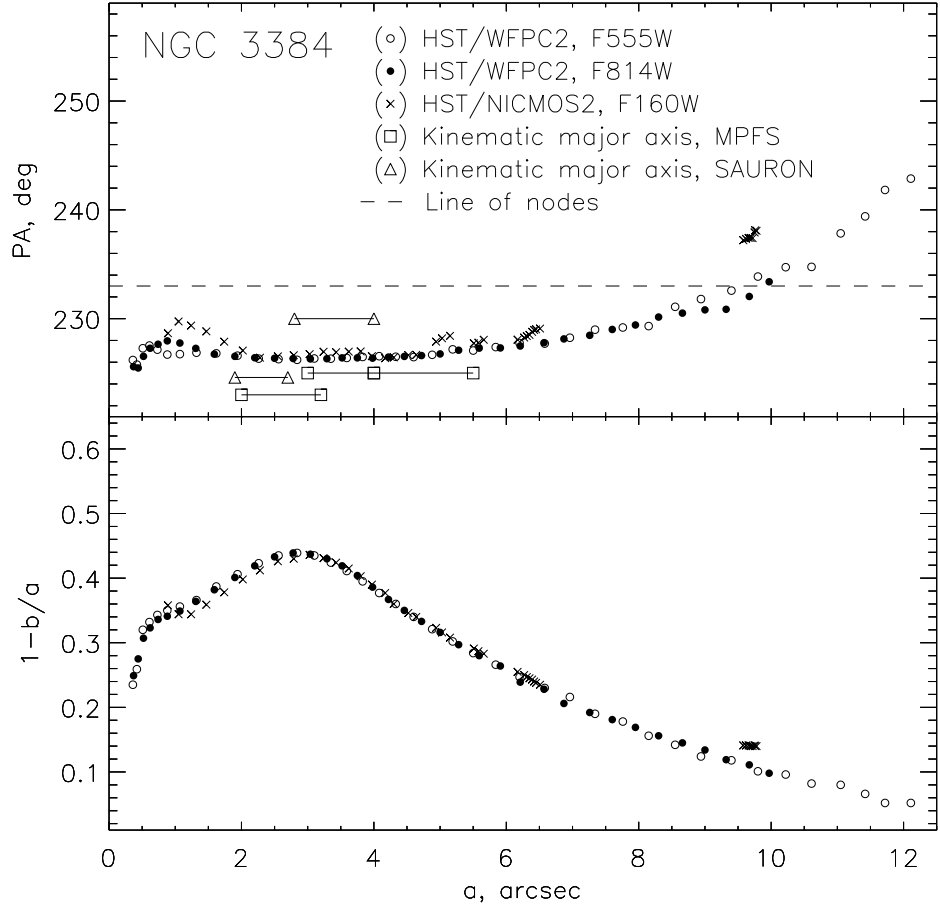}
\caption{Comparison of the orientations of the isophotal
(photometric) major axis and the kinematical major axis (see the text)
in the center of NGC~3384 (top) and circumnuclear radial variations of the
isophotal ellipticity (bottom); at the upper plot an orientation of the
outer disk line-of-nodes is shown by the dashed line}
\end{figure}

\begin{figure}
\epsscale{1}
\plotone{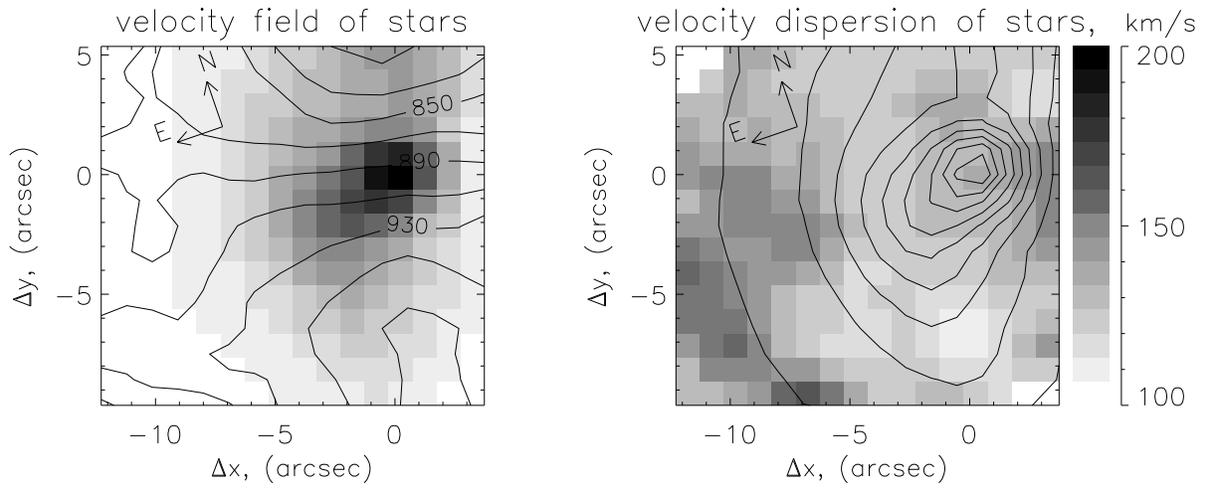}
\caption{Line-of-sight velocity field for the stellar
component (isovelocity lines overlaid onto the continuum image) and stellar
velocity dispersion map (grayscale, with green continuum isophotes
overlaid) in NGC~3368 according to the MPFS data. The images are direct,
$PA(top)=-19^{\circ}$}
\end{figure}

\clearpage

\begin{figure}
\plotone{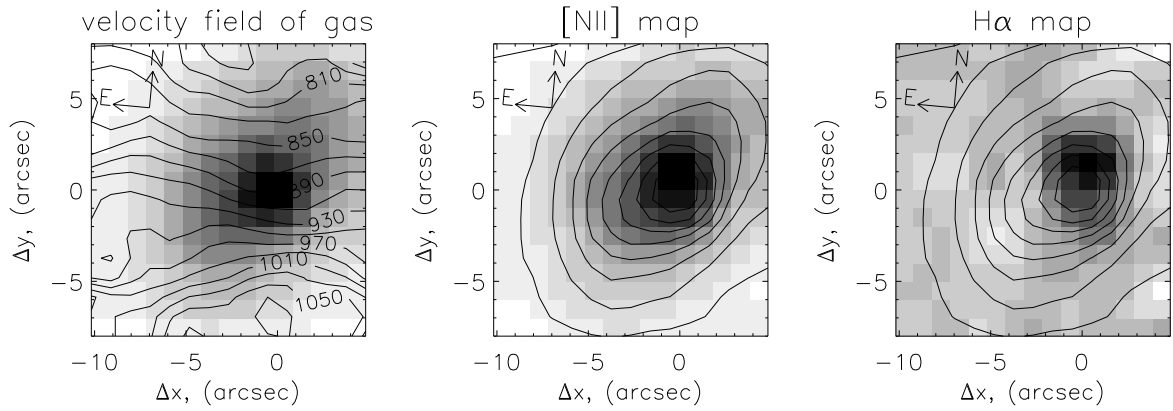}
\caption{Line-of-sight velocity field for the ionized gas
(isovelocity lines overlaid onto the continuum image) and emission-line
surface brightness distributions (grayscale, with red
continuum isophotes overlaid) in NGC~3368 according to the MPFS data.
The images are direct, $PA(top)=+5^{\circ}$}
\end{figure}

\begin{figure}
\epsscale{0.7}
\plotone{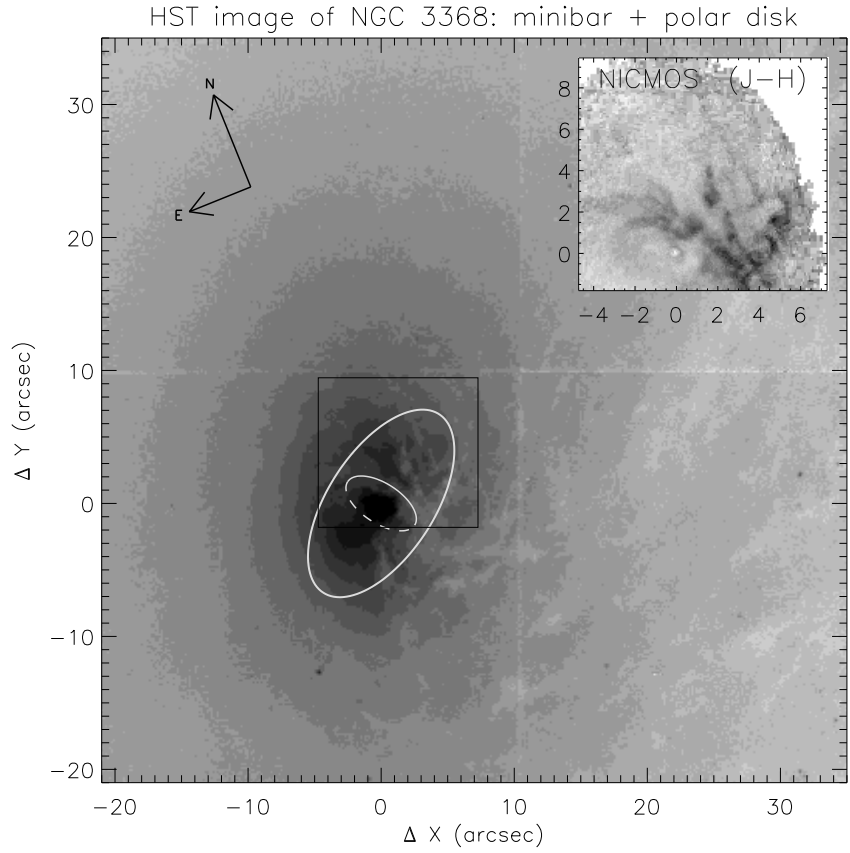}
\caption{An optical-band WFPC2/HST image of NGC 3368 combined from
raw data in several filters in order to best reveal the dust patches;
two ellipses
delineate the orientation of the outermost isophote (the larger one) and the
circumnuclear polar dust ring (the smaller one). For the area outlined
by a square the color map obtained by dividing the NICMOS/F160W image
by the NICMOS/F110W image is shown in the upper right corner; the
darker features are the redder ones}
\end{figure}

\begin{figure}
\epsscale{0.6}
\plotone{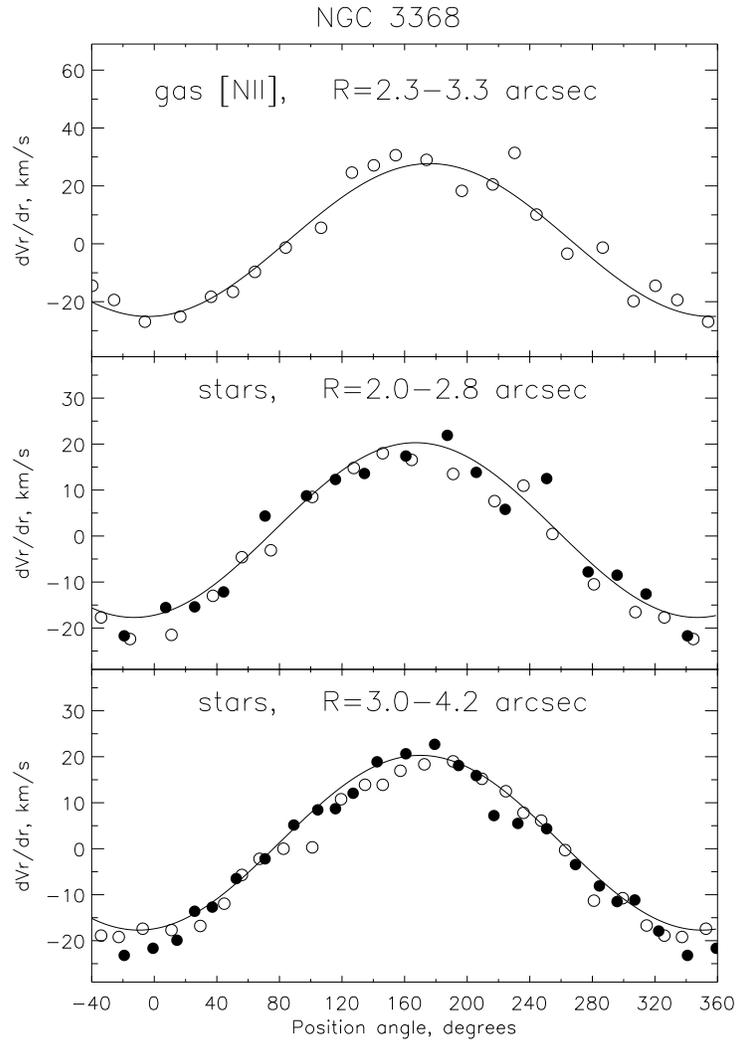}
\caption{Azimuthal dependence of the central line-of-sight
velocity gradient for the ionized gas and stars in the center of NGC~3368
in several radial bins; filled and open circles represent the MPFS
data of different observing dates; solid lines show the best-fit
sinusoids}
\end{figure}

\begin{figure}
\epsscale{0.8}
\plotone{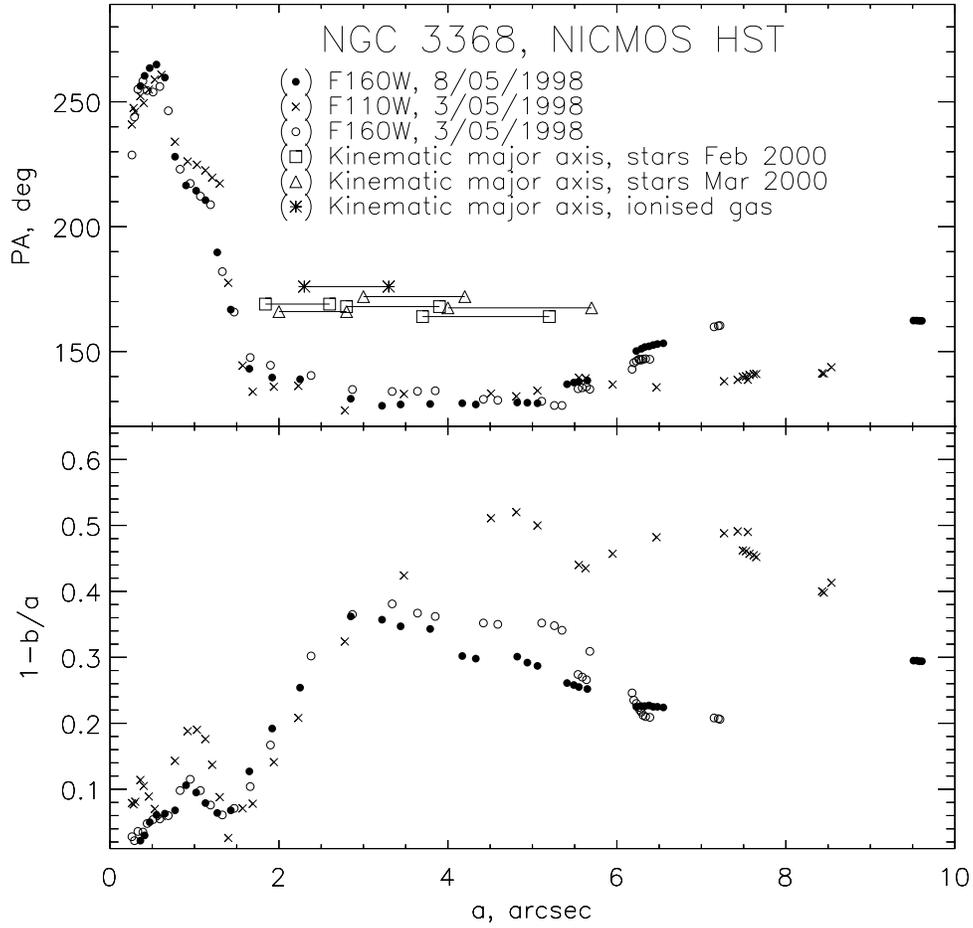}
\caption{Comparison of the orientations of the isophotal
(photometric) major axis and the kinematical major axis (see the text)
in the center of NGC~3368 and circumnuclear radial variations of the
isophotal ellipticity}
\end{figure}

\begin{figure}
\plotone{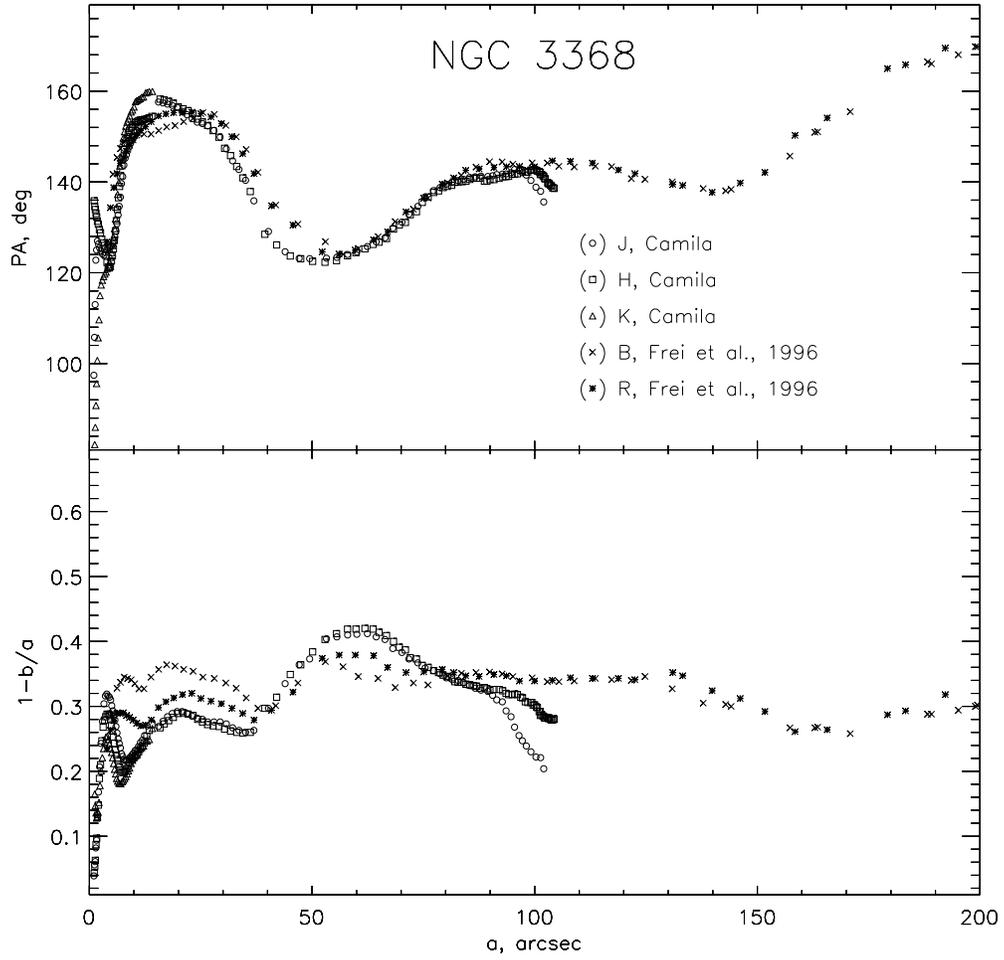}
\caption{Radial variations of the isophotal characteristics in NGC 3368
according to our near-infrared and to Frei et al.'s (1996) optical
photometric data}
\end{figure}

\begin{figure}
\epsscale{1}
\plotone{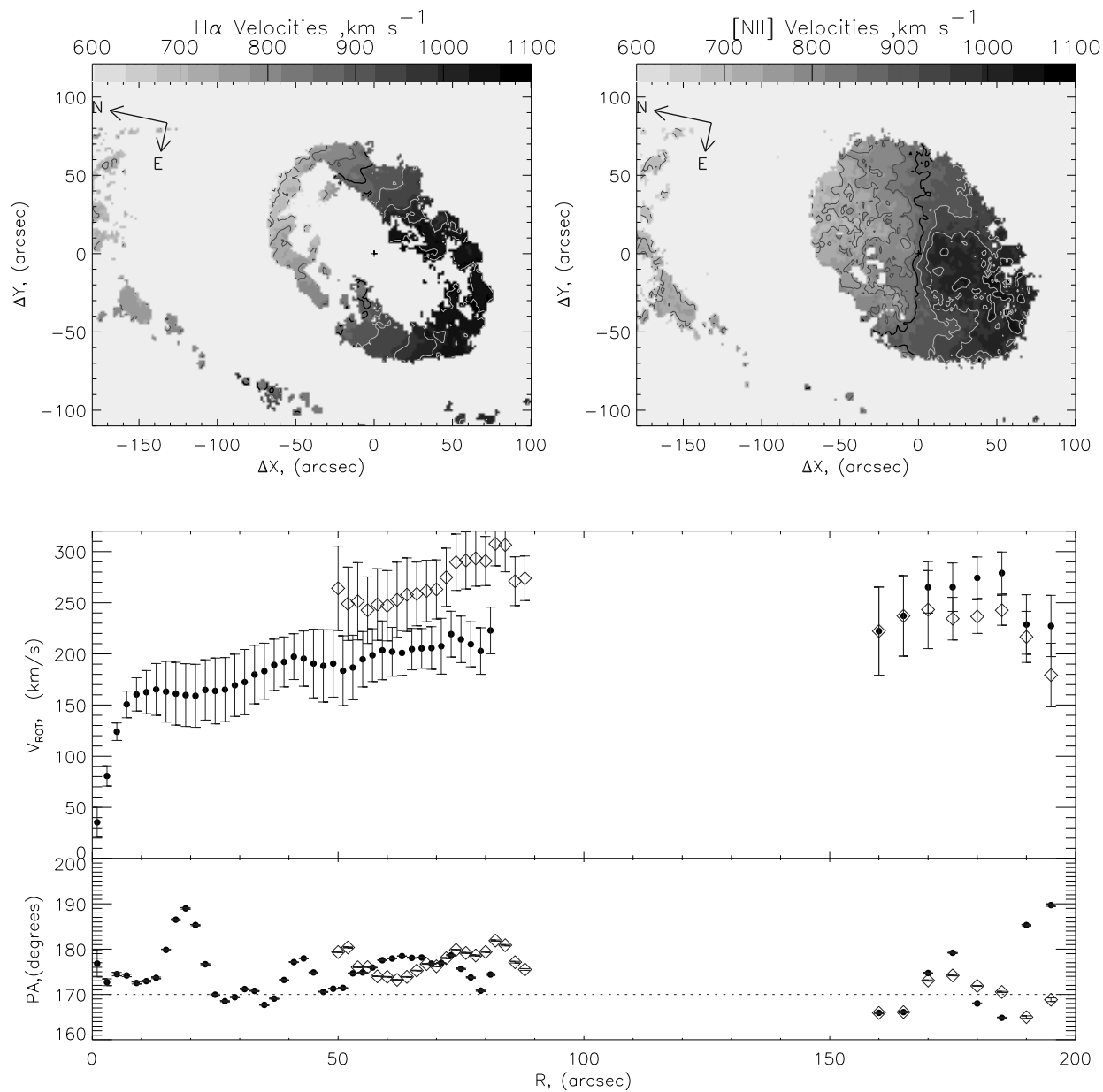}
\caption{Results of the observations of NGC~3368 with the IFP: {\it top}
-- the velocity fields in the H$\alpha$ (left) and [NII] (right) emission
lines. The step between isovelocity contours is
$50\, \mbox{km} \cdot \mbox{s}^{-1}$.
The black contour marks the systemic velocity of
the galaxy center; {\it bottom} -- the radial distributions of the rotational
velocities and $PA$ of the kinematical axis that are calculated on the
basis of the
ionized-gas velocity fields. Open diamonds and filled circles correspond
to the measurements in the H$\alpha$ and [NII] lines, respectively.
The dotted line marks the $PA$ of the gaseous disk (see the text)}
\end{figure}

\end{document}